\documentclass[twoside,twocolumn,9pt]{article}
\usepackage{extsizes}
\usepackage[super,sort&compress,comma]{natbib} 
\usepackage[version=3]{mhchem}
\usepackage[left=1.5cm, right=1.5cm, top=1.785cm, bottom=2.0cm]{geometry}
\usepackage{balance}
\usepackage{times,mathptmx}
\usepackage{sectsty}
\usepackage{graphicx} 
\usepackage{lastpage}
\usepackage[format=plain,justification=justified,singlelinecheck=false,font={stretch=1.125,small,sf},labelfont=bf,labelsep=space]{caption}
\usepackage{float}
\usepackage{fancyhdr}
\usepackage{fnpos}
\usepackage[english]{babel}
\addto{\captionsenglish}{%
  
}
\usepackage{array}
\usepackage{droidsans}
\usepackage{charter}
\usepackage[T1]{fontenc}
\usepackage[usenames,dvipsnames]{xcolor}
\usepackage{setspace}
\usepackage[compact]{titlesec}
\usepackage{hyperref}

\usepackage{subfigure}
\usepackage{breqn}
\usepackage{amssymb}
\usepackage{mhchem}
\usepackage{notes2bib}

\usepackage{epstopdf}

\definecolor{cream}{RGB}{222,217,201}

\begin{document}

\pagestyle{fancy}
\thispagestyle{plain}

\makeFNbottom
\makeatletter
\renewcommand\LARGE{\@setfontsize\LARGE{15pt}{17}}
\renewcommand\Large{\@setfontsize\Large{12pt}{14}}
\renewcommand\large{\@setfontsize\large{10pt}{12}}
\renewcommand\footnotesize{\@setfontsize\footnotesize{7pt}{10}}
\makeatother

\renewcommand{\thefootnote}{\fnsymbol{footnote}}
\renewcommand\footnoterule{\vspace*{1pt}% 
\color{cream}\hrule width 3.5in height 0.4pt \color{black}\vspace*{5pt}} 
\setcounter{secnumdepth}{5}

\makeatletter 
\renewcommand\@biblabel[1]{#1}            
\renewcommand\@makefntext[1]% 
{\noindent\makebox[0pt][r]{\@thefnmark\,}#1}
\makeatother 
\renewcommand{\figurename}{\small{Fig.}~}
\sectionfont{\sffamily\Large}
\subsectionfont{\normalsize}
\subsubsectionfont{\bf}
\setstretch{1.125} %In particular, please do not alter this line.
\setlength{\skip\footins}{0.8cm}
\setlength{\footnotesep}{0.25cm}
\setlength{\jot}{10pt}
\titlespacing*{\section}{0pt}{4pt}{4pt}
\titlespacing*{\subsection}{0pt}{15pt}{1pt}

\fancyhead{}
\renewcommand{\headrulewidth}{0pt} 
\renewcommand{\footrulewidth}{0pt}
\setlength{\arrayrulewidth}{1pt}
\setlength{\columnsep}{6.5mm}
\setlength\bibsep{1pt}
%%%END OF FOOTER%%%

\makeatletter 
\newlength{\figrulesep} 
\setlength{\figrulesep}{0.5\textfloatsep} 

\newcommand{\topfigrule}{\vspace*{-1pt}% 
\noindent{\color{cream}\rule[-\figrulesep]{\columnwidth}{1.5pt}} }

\newcommand{\botfigrule}{\vspace*{-2pt}% 
\noindent{\color{cream}\rule[\figrulesep]{\columnwidth}{1.5pt}} }

\newcommand{\dblfigrule}{\vspace*{-1pt}% 
\noindent{\color{cream}\rule[-\figrulesep]{\textwidth}{1.5pt}} }

\makeatother
\twocolumn[
  \begin{@twocolumnfalse}
\vspace{3cm}
\sffamily

\noindent\LARGE{\textbf{Atomistic simulation of PDADMAC/PSS oligoelectrolyte multilayers: overall comparison of tri- and tetra-layer systems$^\dag$}}
\vspace{0.3cm} \\

\noindent\large{Pedro A. S\'anchez,$^{\ast}$\textit{$^{a,b}$} Martin V\"{o}gele,\textit{$^{c}$} Jens Smiatek,\textit{$^{d}$} Baofu Qiao,\textit{$^{e}$} Marcello Sega,\textit{$^{f}$} and Christian Holm\textit{$^{d}$}} 

\vspace{0.6cm}

\noindent\normalsize{By employing large-scale molecular dynamics simulations of atomistically resolved oligoelectrolytes in aqueous solutions, we study in detail the first four layer-by-layer deposition cycles of an oligoelectrolyte multilayer made of poly(diallyl dimethyl ammonium chloride)/poly(styrene sulfonate sodium salt) (PDADMAC/PSS). The multilayers are grown on a silica substrate in 0.1M {NaCl} electrolyte solutions and the swollen structures are then subsequently exposed to varying added salt concentration. We investigated the microscopic properties of the films, analyzing in detail the differences between three- and four-layers systems. Our simulations provide insights on the early stages of growth of a multilayer, which are particularly challenging for experimental observations. We found a rather strong entanglement of the oligoelectrolytes, with a fuzzy layering of the film structure. The main charge compensation mechanism is for all cases intrinsic, whereas extrinsic compensation is relatively ehanced for the layer of the last deposition cycle. In addition, we quantified other fundamental observables of these systems, as the film thickness, water uptake, and overcharge fractions for each deposition layer.}

 \end{@twocolumnfalse} \vspace{0.6cm}

  ]
\renewcommand*\rmdefault{bch}\normalfont\upshape
\rmfamily
\section*{}
\vspace{-1cm}

\footnotetext{\textit{$^{a}$~ Ural Federal University, 51 Lenin av., Ekaterinburg, 620000, Russian Federation. E-mail: r.p.sanchez@urfu.ru}}
\footnotetext{\textit{$^{b}$~b Institute of Ion Beam Physics and Materials Research, Helmholtz-Zentrum Dresden-Rossendorf e.V., Dresden, Germany. }}
\footnotetext{\textit{$^{c}$~Department of Computer Science, Stanford University, Stanford, California, USA. }}
\footnotetext{\textit{$^{d}$~Institut f\"ur Computerphysik, Universit\"at Stuttgart, 70569 Stuttgart, Germany. }}
\footnotetext{\textit{$^{e}$~Chemical Sciences and Engineering Division, Argonne National Laboratory, Argonne, Illinois, USA. }}
\footnotetext{\textit{$^{f}$~Forschungszentrum J\"ulich, Helmholtz Institute Erlangen-Nuremberg, Nuremberg, Germany. }}

\section{Introduction}
Charged polymers, or polyelectrolytes (PEs), are one of the most relevant types of macromolecules. They include essential biochemical compounds, like DNA and RNA, and numerous synthetic polymer materials. The main common characteristic of these polymers is that their monomers include electrolyte groups that dissociate in water, forming a solution of charged polymer backbones--either polyanions or polycations--and their respective counter-ions. This allows PEs to assemble with oppositely charged substances to form electrostatically stabilized complexes \cite{2004-thuenemann}.

The electrostatic assembly properties of PEs is the basis of an important synthesis technique that provides layered hybrid polymer thin films. In this experimental procedure, known as layer-by-layer deposition (LbL) \cite{1992-decher, decher97a}, a substrate with a surface charge is exposed to a solution of PEs with opposite charge, typically by simply dipping the substrate wafer into the PE solution. During this dipping process, part of the PEs can be electrostatically adsorbed on the substrate surface, forming a first film layer. Non adsorbed PEs and counter-ions are then removed, usually by rinsing the wafer with pure solvent. A second polymer layer can be deposited on top of the first one by repeating the dipping process with a solution of oppositely charged PEs. In this way, large amounts of layers can be deposited by alternating the exposition of the wafer to solutions of polyanions and polycations. The films created with this technique, known as polyelectrolyte multilayers (PEMs), can be used in many technological applications \cite{2007-ariga, 2009-schlenoff, 2012-decher}. For instance, PEMs materials have been used for the creation of membranes in filtration and catalysis systems \cite{2005-malaisamy, 2008-bruening, 2008-datta, 2015-degrooth-jms, 2015-degrooth-jms2}, antimicrobial and biocompatible protective coatings \cite{2003-thierry, 2015-seon-lgm}, tissue engineering and single cell analysis systems \cite{2012-gribova, 2014-volodkin-pol}, nanocapsules and responsive coatings for drug delivery and dose control applications \cite{2005-khopade, 2016-micciulla-sm}, as matrix materials for active components in solar cells or biosensor applications\cite{2003-trau}, as well as elements of non-linear optical materials \cite{2005-arsenault, 2005-jiang}. During the last two decades the great interest of PEMs based materials has stimulated the development of different experimental enhancements to the LbL deposition technique \cite{1999-dubas, 2001-cho, 2001-chiarelli, 2000-schlenoff, 2009-kolasinska}, together with large experimental and theoretical research efforts to achieve a proper understanding of the physical mechanisms behind the multilayer growth process \cite{2003-schoenhoff, 2004-klitzing, 2007-schoenhoff, 2009-cerda-eur}. For example, numerous works have been devoted to the characterization of the different growth regimes that can be obtained from the LbL deposition \cite{2001-mcaloney, 2002-lavalle, 2002-picart, 2003-mcaloney, 2003-schoeler, 2004-lavalle, 2009-hoda, 2014-volodkin, 2016-tang-sm, 2016-vikulina-pccp, 2017-bellanger-jcis}. The dependence on the charge density of the PEs in the formation and stability of the multilayers has been also a subject largely studied \cite{2002-glinel, 2003-schoeler, 2004-mueller, 2007-gopinathan-prl, 2008-salomaeki, 2009-haitami}. Several other aspects determining the formation and properties of PEMs, as the nature of the ions in the dipping solution \cite{2001-dubas, 2004-salomaki, 2007-gopinadhan-jcpb, 2009-haitami}, the internal distributions of ions and solvent \cite{2002-riegler, 2008-tanchak, 2008-garg} or the short-range interactions,\cite{2004-klitzing} have been also addressed. More detailed reviews on available studies on PEMs systems can be found in references\cite{2003-schoenhoff, 2009-cerda-eur, 2012-cerda}. 

Despite the large research efforts already made on PEMs systems, their proper understanding is still far from complete. The complexity of the interactions during the growth process and the strong structural correlations adopted by oppositely charged PEs produce a significant intermixing and complexation between different layers. Since the resolution power of current experimental techniques only provides measurements at length scales not smaller than several tens of nanometers \cite{2009-block, 2010-roiter}, direct observations of the internal nanostructure of PEMs systems are still out of reach. Available analytical models have difficulties to deal properly with large structural correlations and often rely on strong unproven assumptions as, for example, the frequently assumed equilibrium nature of the multilayer structure \cite{2002-kovacevic}. Furthermore, the computer simulation of PEMs is also challenging. To date, a main part of computational studies has been based on coarse-grained top-down modeling approaches with very crude approximations \cite{2009-cerda-sm, 2010-reddy}. The relative low computing cost of coarse-grained models allows the theoretical study of different properties of PEMs at mesoscopic length and times scales, easing the comparison to experimental measurements \cite{2004-messina-jpsb, 2004-messina-mm, 2005-patel, 2006-patel, 2015-narambuena}. However, the top-down strategy has difficulties to accurately connect such experimentally observable mesoscopic scales to the actual microscopic properties of the films. On the other hand, even though atomistic simulations have the possibility to provide insights at the microscopic scale, their high computing cost imposes strong limitations on the time and length scales that can be reached.

A way to reduce the computing cost of fully atomistic simulations is to focus on modeling of systems of PE oligomers, or oligoelectrolytes (OEs). In this way, one can decrease significantly the amount of atoms to be simulated, reaching scales that allow a direct comparison with experimental measurements. This provides direct insights on microscopic properties and physical mechanisms that might be also relevant in systems of larger PEs, paving the way for the bottom-up design of accurate coarse-grained models \cite{2015-voegele}. Nevertheless, the study of specific properties of OEs complexes and multilayers is an important topic by itself. Oligoelectrolyte multilayers (OEMs) are expected to have a different structure and faster dynamics than conventional, high molecular weight PEMs, while keeping the same surface chemistry. OEMs would thus show different porosity and mechanical properties, as well as a higher sensitivity to external stimuli and a faster degradation, which can be an advantage for specific applications. Some of such differences have been confirmed in recent experimental measurements, showing a stronger dependence of the stability of OEMs on the preparation conditions, as well as a weaker variation of the elasticity of the film with the number of deposited layers \cite{2014-micciulla-pccp}.

Inspired by the considerations discussed above, we introduced several pioneering works on the fully atomistic simulation of OEs complexes \cite{2010-qiao}, monolayers \cite{2011-qiao-mm} and bilayers \cite{2011-qiao-pccp, 2012-qiao-pccp}. The latter were obtained by following a simulation protocol that mimics the experimental LbL deposition method. A subsequent extension to a tetralayer of short poly(diallyl dimethyl ammonium chloride)/poly(styrene sulfonate sodium salt (PDADMAC/PSS) chains deposited on a charged silica substrate allowed us to make the first direct comparison of a set of mesoscopic properties of an OEM system---film thickness, surface roughness and amount of adsorbed chains---measured in atomistic simulations and experiments \cite{2014-micciulla-softmat}. The good agreement between the results obtained from both approaches suggests that the underlying microscopic mechanisms are well captured by our simulations. Since the comparison of mesoscopic properties was the main goal in such work, detailed microscopic properties of the system that could not be directly measured in experiments were not discussed there. However, is at such microscopic scale where atomistic simulations excel, being able to provide fundamental understanding of larger scale phenomena. Therefore, this microscopic information can be considered the main outcome of our simulations, and is the main subject of this work: here, we present the microscopic details of the internal structure and charge properties of the simulated PDADMAC/PSS OEM, comparing the three (3L) and four (4L) layers systems. We also introduce a comparison of the properties of these systems when shortly exposed to solutions with different added salt concentrations. This does not correspond to salt annealing processes, as the time scales involved in the latter are out of reach for atomistic simulations, but to a post-deposition short rinsing. In order to better analyze the changes experienced by the OE chains due to their multilayer arrangement, we also performed simulations of single PDADMAC and PSS chains in bulk under the same concentrations of added salt, that will serve as reference when discussing diverse properties of the multilayers.

The paper is organized as follows: first, we describe the details of the system under study and the protocols used for the simulation of both, the OE tetralayer and the corresponding OE single chains in bulk; then, we start the discussion of the results with an analysis of the general film structure that follows an increasing level of detail, from the overall film density profiles to the single OE chain structure; next, we focus on charge properties, pair correlation functions and overcharging effects; finally, we summarize the main conclusions and provide an outlook.

\section{System and simulation details}
PDADMAC and PSS are a complementary pair of polyelectrolytes frequently used in the synthesis of PE complexes and multilayers. They are strong PEs---\textit{i.e.}, they become fully charged in solution---for a broad range of pH values and low added salt concentrations, including the range sampled here for the latter. This simplifies the simulation of the effects of different ionic strengths on their structure.

\subsection{PDADMAC/PSS OE tetralayer}
The details of the experimental system under study and the approach used for its atomistic simulation were already introduced in references \cite{2014-micciulla-softmat, 2015-sanchez-hlrs}. Briefly, we simulate the growth of a PDADMAC/PSS tetralayer by LbL deposition from water solutions of oligoelectrolytes with a degree of polymerization of $DP=30$ monomers and a concentration of 0.1M of monovalent added salt. The use of added salt has been proven experimentally to be essential for the formation of a stable multilayer when so short PE chains are employed \cite{2014-micciulla-pccp}. Both, counter-ions and added salt ions, are Na$^+$ and Cl$^-$. The adsorbing substrate is a flat silica wafer. This material shows a negative surface charge at intermediate pH values \cite{2001-behrens, 2002-shin}. Therefore, the first and third layers correspond to the adsorption of PDADMAC chains and the second and fourth to PSS ones.

\begin{figure}[!t]
\centering
\includegraphics[width=8.2cm]{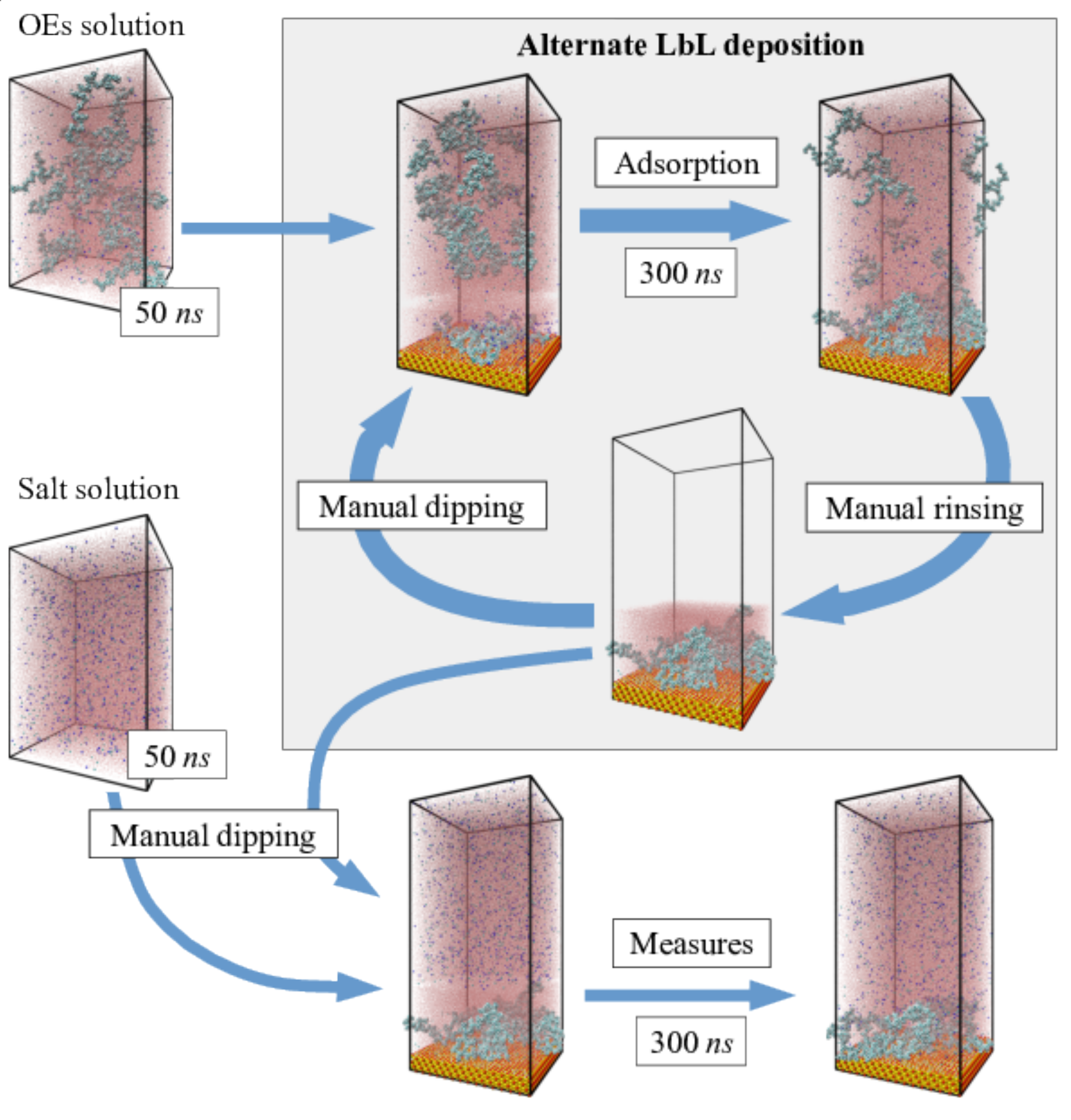}
\caption{Scheme of the protocol used for the simulation of the LbL growth of a PDADMAC/PSS OE tetralayer and its exposition to solutions with different ionic strength.}
\label{fig:simprotocol}
\end{figure}

For these simulations, a model flat silica wafer was created as a four layers silicate sheet with lateral periodic structure of length 13.1650~nm and 12.1608~nm in the $x$ and $y$ directions, respectively. Its surface was made hydrophilic by replacing the uppermost oxygen atoms in every silicate tetrahedron with polar hydroxyl groups (\ce{-OH}). Such hydroxyl groups were allowed to freely rotate with respect to the axis normal to the surface in order to avoid the simulation artifacts known to appear when their orientation is fixed \cite{2011-ho, 2011-qiao-pccp}. The surface charge of the substrate was created by skipping one in every four of the allocations of hydroxyl groups, leaving the corresponding SiO$_2^-$ groups exposed. This leads to an average surface charge of $\sigma=-0.03$ C/m$^2$, in good agreement with experimental values \cite{2001-behrens, 2002-shin}. This simple approach assumes that the local microscopic distribution of charges has no relevance on the mesoscopic properties of the system, as has been shown in recent computational studies of electro-osmotic flows within charged walls \cite{2009-smiatek}.

The main steps of the simulation protocol, designed to reproduce the basic features of the LbL growth process, are sketched in Figure~\ref{fig:simprotocol}. This protocol includes several direct manipulations of the system (labeled as `Manual dipping' and `Manual rinsing' in Figure~\ref{fig:simprotocol}) and different sequential simulation runs (`OEs solution', `Salt solution', `Adsorption' and `Measures' in Figure~\ref{fig:simprotocol}). In all of such simulation runs, performed with the simulation package Gromacs 4.5.3\cite{2008-hess}, the OPLS-AA force field \cite{1996-jorgensen} was used to represent the interactions of the OEs. Details of the parametrization of this force field to the system under study can be found in reference \cite{2010-qiao}. The solvent was modeled by means of the SPC/E water model, that is able to reproduce accurately its dielectric constant \cite{1987-berendsen, 2006-hess-prl, 2006-hess-jcp}, in combination with the SETTLE algorithm \cite{1992-miyamoto} when imposing molecular geometry constrains. Van der Waals forces were represented as Lennard-Jones potentials shifted to smoothly decay to zero at distances larger than 0.9~nm. Electrostatic interactions were calculated using the particle mesh Ewald method (PME) \cite{1993-darden, 1995-essmann} with a Fourier spacing of 0.125~nm and a direct space cutoff of 1.3~nm. We used a rectangular simulation box with lateral periodic boundaries in the $x$ and $y$ directions, with the adsorbing substrate placed at the bottom and parallel to the $x$--$y$ plane. In order to take into account the slab geometry of the system in the calculation of the electrostatic interactions, the Yeh-Berkowitz correction \cite{1999-yeh} was applied using an upper empty region of the simulation box of twice the height of its filled region.

The first step in each LbL cycle is the dipping of the adsorbing substrate into alternating solutions of PDADMAC and PSS chains. The simulated samples corresponding to such solutions were obtained from the solvation of 20 OE chains, their respective 600 counter-ions (Cl$^-$ for PDADMAC and Na$^+$ for PSS) and 192 Na$^+$ and Cl$^-$ ions of added salt, in 99286 water molecules. These systems were first prepared using the package Packmol 1.1.1\cite{2009-martinez} in a simulation box with the same lateral dimensions of the silica substrate and then equilibrated for 50~ns in separate molecular dynamics (MD) simulations in the NPT ensemble (labeled as `OEs solution' in Figure~\ref{fig:simprotocol}), imposing a pressure of 1~bar in the $z$ direction by means of a semi-isotropic Parrinello-Rahman barostat \cite{1981-parrinello, 1983-nose} and a temperature of $T=$298~K using a Nos\'{e}-Hoover thermostat with a relaxation time of $\tau=$0.5~ps. The dipping of the deposition wafer into the OEs solutions was mimicked by simply placing the former in the bottom of the simulation box and the latter on top of it (`Manual dipping' in Figure~\ref{fig:simprotocol}). This combined system was then allowed to relax in a simulation slightly longer than 300~ns, in which a fraction of the OE chains from the dipping solution were adsorbed on the substrate and the layers deposited in previous cycles (`Adsorption' in Figure~\ref{fig:simprotocol}). In fact, the simulation of the OEs adsorption process, that is the crucial part in our protocol, was performed in several steps: first, an initial relaxation of the manually assembled system was performed by energy minimization with the steepest descent algorithm, followed by three sequential MD runs in the NVT ensemble with the same thermostat and reference temperature mentioned above; the first of such runs consisted of a 100~ps MD simulation with an integration time step of $\delta t=1$~fs; the second run, also of 100~ps with a time step of $\delta t=2$~fs, where the covalent bonds of the hydrogen atoms were constrained with the LINCS algorithm\cite{1997-hess}; finally, a production run of 300~ns with $\delta t=2.5$~fs where all covalent bonds were constrained with the LINCS algorithm. During these simulations, the amount of adsorbed OE chains was calculated by counting the number of charged groups in close contact with the substrate or other adsorbed OE chains. where closeness was defined to be nearer than 1 nm. As a result, we checked that a stationary value for the amount of adsorbed chains was reached within the simulation time in all cases.

Each LbL growth cycle ends with the rinsing of the film wafer with pure solvent or a pure salt solution in order to remove the supernatant PEs. Here we mimic this procedure by simply removing from the simulation box all non-adsorbed OEs, all added salt ions and all water molecules placed at a distance from the substrate larger than the uppermost atom of adsorbed OE chains (`Manual rinsing' in Figure~\ref{fig:simprotocol}). The resulting rinsed system was then used as the starting point for the next LbL cycle. Finally, the rinsed systems obtained after the third and fourth LbL cycles were exposed to three solutions with different salt concentrations, $c$=0 (pure solvent), 0.1M and 0.5M. This was done by using the same `manual dipping' procedure described above, with one of such solutions replacing the solution of OEs. Measures of the system properties were taken during the last 100~ns of a subsequent simulation of 300~ns with the same steps described above for the OEs adsorption part. In order to estimate the statistical variance of such measurements, three independent runs were performed for each system.

\subsection{PDADMAC/PSS single OE chains}
In order to analyze the changes in the structural properties of individual OE chains that become part of an OEM structure, we performed MD atomistic simulations of single short PDADMAC and PSS chains in presence of the same three added NaCl salt concentrations sampled for the tetralayer system, $c=0$, 0.1~M and 0.5~M. In this case we varied the degree of polymerization, $DP$, from 10 to 30 in steps of 5. These simulations were also performed using the OPLS-AA force field and SPC/E water model. Each simulated system consisted of one OE chain solvated in a cubic box of 8~nm of side length and the number of NaCl ions corresponding to OE counter-ions and chosen added salt concentration.

The simulations involved three equilibration and one production consecutive steps. The first equilibration step was an energy minimization using a steepest descent algorithm for up to 1000 integrations. The second step was a MD simulation of 2~ns in the NVT ensemble, using an integration time step of $\delta t=1$~fs and keeping a temperature of $T=298$~K by means of a Berendsen thermostat with relaxation time of $\tau=0.5$~ps. The third equilibration step was a NPT MD simulation of 10~ns with the same $\delta t$, using a Nos\'{e}-Hoover thermostat with the same reference temperature and relaxation time, together with an isotropic Parrinello-Rahman barostat at a reference pressure of 1~bar and relaxation time of 4~ps. Finally, the production run was an extension of the latter step, using an increased integration time step of $\delta t=2$~fs. The total time of this production run depended on the length of the OE chain: 200~ns for $DP=\{10, 15\}$, 300~ns for $DP = \{20, 25\}$ and 400~ns for $DP = 30$.

For these systems, Gromacs 4.6.7\cite{2008-hess} was used for the MD simulations and the MDAnalysis Python library\cite{2011-michaud-agrawal} for the analysis of the simulation trajectories. 

\section{Results and discussion}

The ionic strength of the background solution\cite{2018-oneal} and the type of OEs forming the outermost layer\cite{2013-ghostine, 2015-zerball} have a strong impact on many properties of technological interest of PDADMAC/PSS PEMs, like their internal structure, porosity, elastic properties, thickness and amount of water retained within the film. For this reason, here we compare the results obtained from our simulations for both, the different sampled concentrations of added salt and the cases of PDADMAC or PSS terminated OEMs. The latter will be mainly done by comparing the trilayer (3L) and the tetralayer (4L) systems. For the sake of clarity, the discussion is presented by following an increasing level of detail: We first discuss the general structure of the simulated films, from overall density profiles and water uptake to the configurations of single OE chains; second, we analyze the charge properties of the OEMs and the growth regime; finally, we discuss atom pairings.

\subsection{General film structure}
\begin{figure*}[!t]
\centering
\subfigure[1L]{\includegraphics[width=4.5cm]{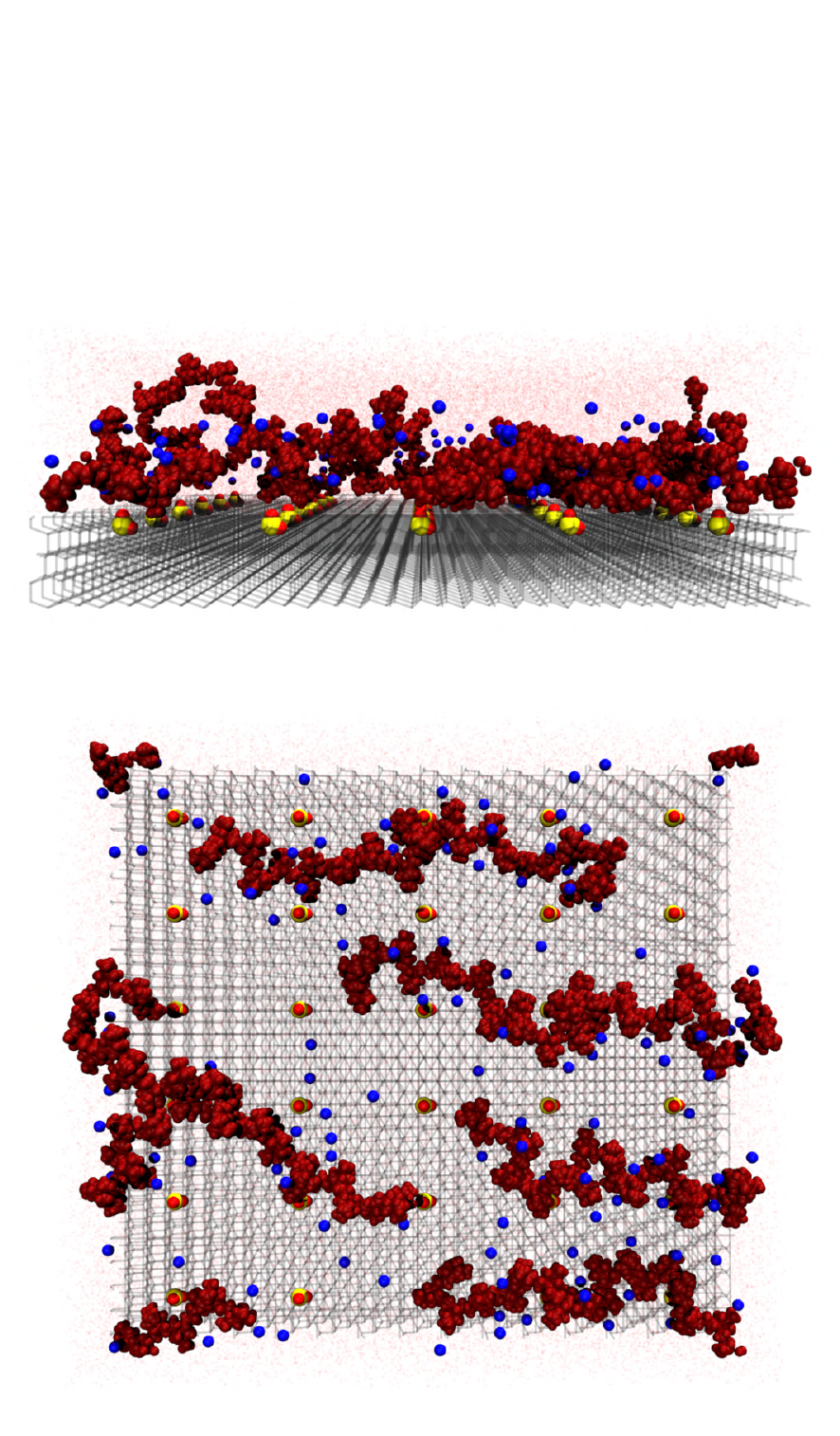}}
\subfigure[2L]{\includegraphics[width=4.5cm]{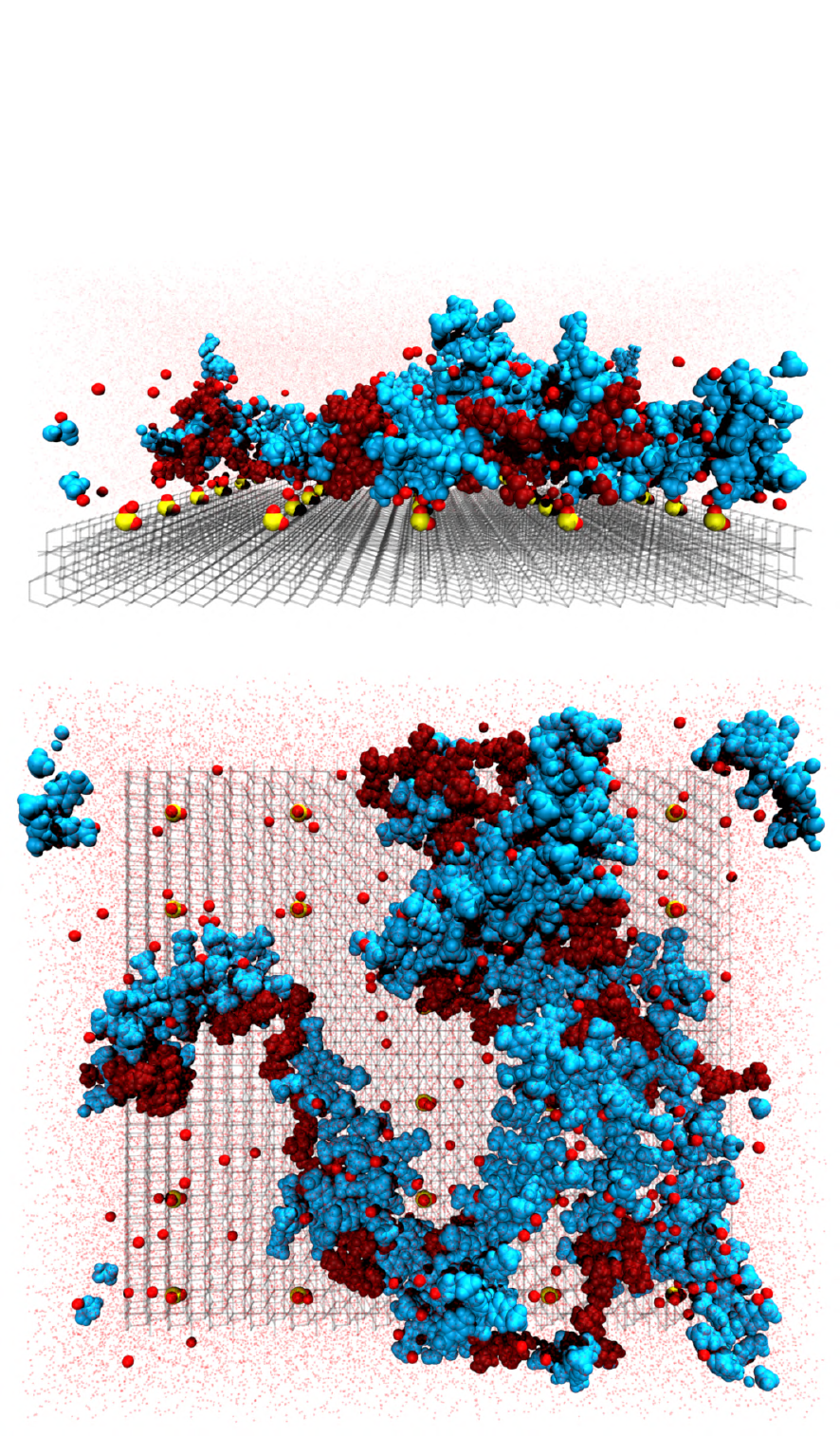}}
\subfigure[3L]{\includegraphics[width=4.5cm]{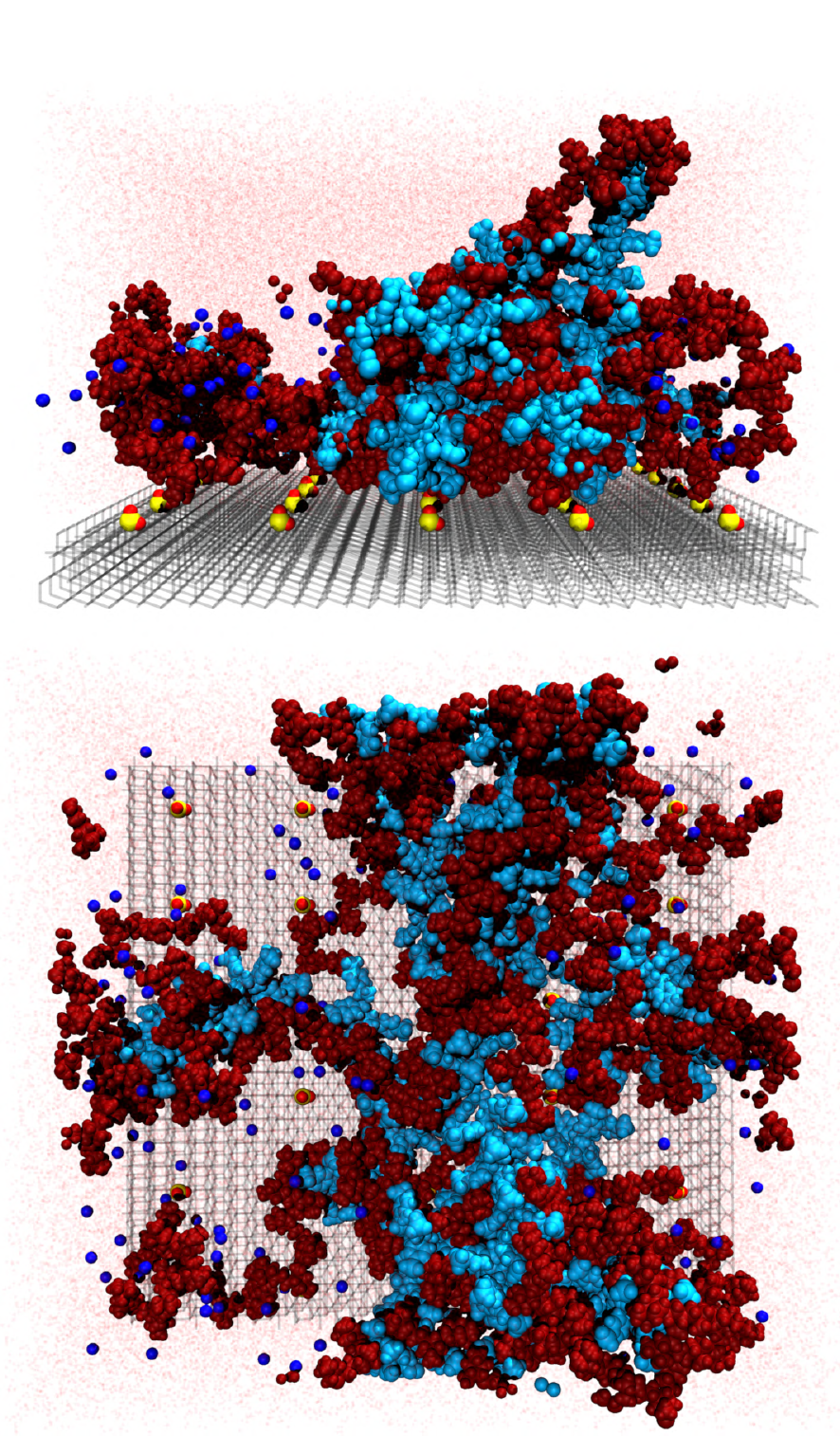}}
\subfigure[4L]{\includegraphics[width=4.5cm]{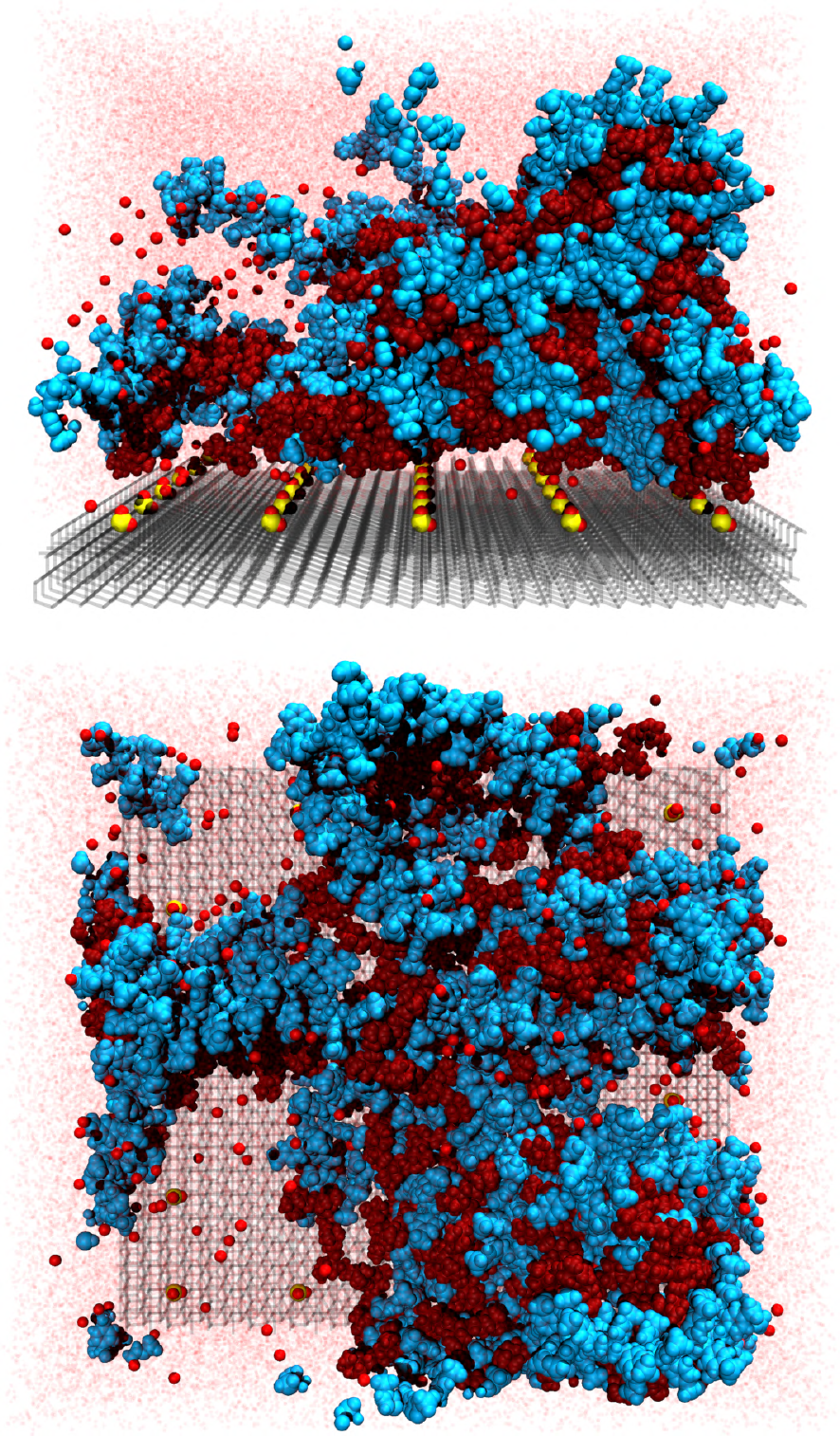}}
\caption{Side view (upper row) and top view (lower row) of simulation snapshots of a growing PDADMAC/PSS multilayer as the number of deposited cycles increases from one to four (columns from left to right). To ease the visualization, PDADMAC chains are colored in dark red and PSS ones in light blue, Na$^+$ ions in light red and Cl$^-$ in dark blue, whereas the silica subtrate is represented as a wireframe except for the charged groups. Configurations correspond to systems obtained after the rinsing procedure.}
\label{fig:snapshots}
\end{figure*}

The simple direct inspection of simulation snapshots is useful to obtain a first qualitative picture of the behavior of the system. Figure \ref{fig:snapshots} shows a set of configuration snapshots that illustrate the growth process of the multilayer. They correspond to configurations obtained from the same independent run right after each deposition and rinsing cycles were performed. As the amount of deposited layers increases from one to four, one can observe how the substrate coverage also tends to grow. However, even after the fourth layer is deposited, a fraction of the substrate surface remains exposed. By comparing 1L and 2L configurations, one can observe that a significant rearrangement of the chains deposited during the first cycle takes place on the second one, with the onset of OEs complexation. After that, single chain rearrangements are difficult to trace but the structure of the regions occupied by OEs seems to experience mainly a simple thickening. Importantly, even though the snapshots are colored to ease the discrimination of PDADMAC and PSS chains, no clear alternating vertical arrangement of layers of polyanions and polycations can be distinguished. Instead, chains of both OEs look very entangled within the film structure.

In the next sections we perform a detailed analysis of the structure of the simulated films by means of different parameters.

\subsubsection{Overall density profiles and water uptake}

\begin{figure}[!h]
\centering
\includegraphics[width=8.2cm]{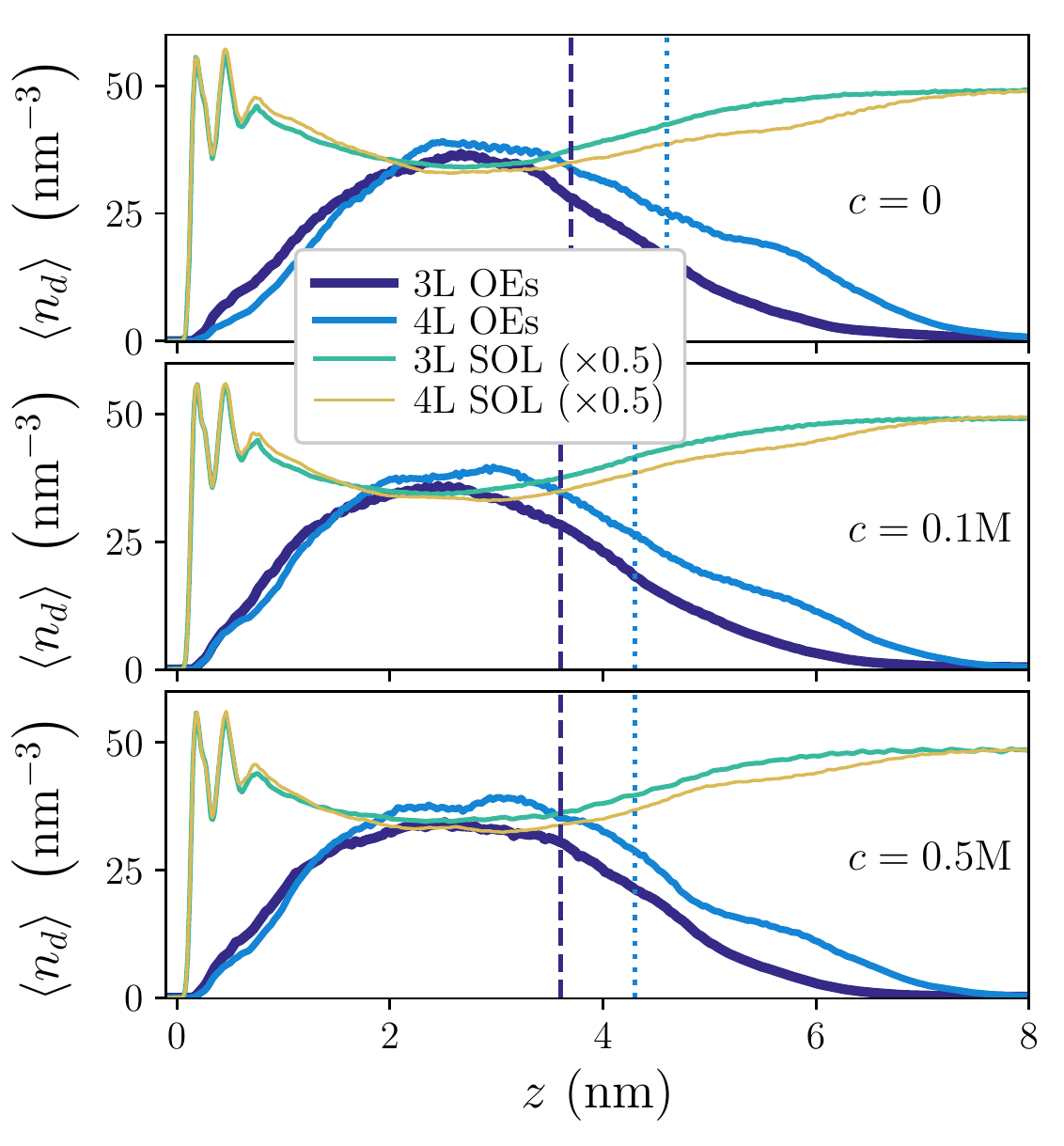}
\caption{Overall oligoelectrolyte (OEs) and water (SOL) atom number density profiles for 3L and 4L systems, averaged over independent runs. Vertical dashed and dotted lines indicate the estimated film thickness, $\langle h^* \rangle$, for 3L and 4L, respectively.}
\label{fig:ovalldensprof}
\end{figure}

The overall density profiles provide a general overview of the spatial distribution of the different species in the system. Figure~\ref{fig:ovalldensprof} shows the average number density profiles, $\langle n_d \rangle$, corresponding to atoms belonging either to OE chains or solvent, as a function of the distance to the substrate surface, $z$, for 3L and 4L systems under each added salt concentration. Independently from $c$, these profiles evidence that OE chains do not reach distances to the substrate longer than 7 (3L) or 8 nm (4L). They also confirm the observation that most part of OE atoms are located in the central region of the film, approximately between 2 and 4 nm from the substrate, with a fast decay of the profile when approaching the substrate and a slower one when approaching the film free surface. The distinction between three different regions, \textit{i.e.}, innermost (close to the substrate), central and outermost (corresponding to the film free surface) is frequently used in PEMs studies\cite{decher97a, 2000-ladam-lm, 2006-porcel-lm}.

Similarly to what was observed for the PSS/PDADMAC bilayer system,\cite{2011-qiao-pccp} water profiles in Figure~\ref{fig:ovalldensprof} show a significant layering near the substrate, with two prominent peaks at distances around $z\approx0.18$~nm and $z\approx 0.46$~nm from the substrate. These maxima correspond to direct interactions of water hydrogen atoms with the silica substrate and to the second hydration shell, respectively, and reveal a strong hydration of the substrate surface, even in presence of OEs. By comparing 3L and 4L, it is seen that the main expansion of the OE profiles as a consequence of the added chains adsorbed during the 4L deposition is found in the outermost region of the film. The central and inner regions, instead, show only small changes corresponding to a slight shift of the 4L film structure away from the substrate. In general, water shows a profile antithetic to the one of OEs, with a significantly lower density in the central region of the film, as has been established in several experimental works\cite{2000-steitz, 2007-nazaran, 2014-micciulla-pccp}. Consequently, the sloppy structure of the multilayer can be attributed to the strong hydration of the silica surface, which hinders the adsorption of OEs. Differences between 3L and 4L water profiles are only significant in the outermost region: there, the 3L system (PDADMAC-ended) has slightly more water than the 4L one (PSS-ended), associated to a slower decay of the OEs profile.

The existence of an alternating variation of the film thickness and degree of swelling of the outermost layer, as the amount of deposited layers increases, has been reported in different experimental works on PEMs\cite{2007-schoenhoff, 2010-zsombor, 2010-iturriramos, 2015-degrooth-jms, 2016-zerball}. For PDADMAC/PSS PEMs, such behavior---known as `even-odd' effect---corresponds to a larger water uptake and relative film thickness in PDADMAC ended multilayers. However, even-odd effects only become significant after a minimum amount of layers, typically above 6, have been deposited. Therefore, the difference in the water uptake of the outermost region in 3L and 4L systems can not be considered a manifestation of such effect. Instead, it agrees qualitatively with experimental observations of a moderate monotonous decrease in water content of PDADMAC/PSS PEMs, grown under low added salt conditions, as the amount of layers increases from 2 to 8.\cite{2010-iturriramos}

A quantification of the difference in water uptake intended to make a comparison with experimental measurements is rather difficult to obtain from such small samples. The first issue is to define an estimator of the film thickness. Here we take the values estimated in our previous works~\cite{2014-micciulla-softmat, 2015-sanchez-hlrs}, which for $c=0$ provided a reasonably good agreement with direct ellipsometry measurements\cite{2014-micciulla-softmat}. This estimation calculates the film thickness, $\langle h^* \rangle$, as the average of the free surface of the film that is defined by the highest positions of OE atoms in a discretized plane parallel to the substrate. These values are plotted in Figure~\ref{fig:ovalldensprof} and included in Table~\ref{tab:wuptake}. Once $\langle h^* \rangle$ has been determined, we can calculate the amount of water molecules contained between the substrate and the film surface as a number density, $\langle N_{\mathrm{SOL}}^*\rangle$, by integrating the water atom number density profile, $n_d^{\mathrm{(SOL)}}$, as
\begin{equation}
 \langle N_{\mathrm{SOL}}^*\rangle=\frac{1}{3 \langle h^* \rangle}\int_0^{\langle h^* \rangle} \left \langle n_d^{(\mathrm{SOL})} \right \rangle dz.
\end{equation}
Results of this calculation for each system are also shown in Table~\ref{tab:wuptake}, next to their corresponding values of $\langle h^* \rangle$. As expected, $\langle h^* \rangle$ is slightly larger for 4L systems than for 3L. Unfortunately, the statistics is too poor to clearly determine the dependence on added salt concentration, but the average values suggest that the thickness of the film in pure water is slightly larger than that corresponding to 0.1 and 0.5M of added salt. This agrees with the experimental observation of a moderate decrease in the thickness of PDADMAC/PSS PEMs when changing their exposure from zero to low concentrations of different added monovalent salt ions, including Na$^+$/Cl$^-$\cite{2001-dubas, 2018-oneal}
\begin{table}[!t]
\small
\centering
\caption{Average film thickness, $\langle h^* \rangle$ (in $nm$), and water uptake, $\langle N^*_{\mathrm{SOL}}\rangle$ (in $nm^{-3}$), measured for each sampled system. Intervals correspond to standard deviations.}
\label{tab:wuptake}
\begin{tabular*}{0.48\textwidth}{@{\extracolsep{\fill}}c c cc c cc}
\hline
   & & \multicolumn{2}{c}{3L}       & & \multicolumn{2}{c}{4L}      \\
   \hline
   $c$ (M)  & & $\langle h^* \rangle$    & $\langle N_{\mathrm{SOL}}^*\rangle$  & & $\langle h^* \rangle$    & $\langle N_{\mathrm{SOL}}^*\rangle$  \\
   \hline
   \begin{tabular}[c]{@{}c@{}}0\\ 0.1\\ 0.5\end{tabular} & & \begin{tabular}[c]{@{}c@{}}3.7$\pm$0.3\\ 3.6$\pm$0.4\\ 3.6$\pm$0.7\end{tabular} & \begin{tabular}[c]{@{}c@{}}25$\pm$2\\ 25$\pm$2\\ 24$\pm$3\end{tabular} & & \begin{tabular}[c]{@{}c@{}}4.6$\pm$0.2\\ 4.3$\pm$0.2\\ 4.3$\pm$0.3\end{tabular} & \begin{tabular}[c]{@{}c@{}}24$\pm$1\\ 24$\pm$1\\ 23$\pm$2\end{tabular}\\
   \hline
   \end{tabular*}
\end{table}
Regarding the estimation of water uptake, the propagation of the uncertainty in $\langle h^* \rangle$ also prevents us from making a strong point, but the trend for average values suggests a slight decrease in 4L with respect to 3L of around 4\% in all cases. Note that the decrease measured experimentally in Reference~\cite{2010-iturriramos} for analogous PEM systems is not larger than 10\%.

\subsubsection{Split density profiles}

\begin{figure}[!t]
\centering
\subfigure[]{\label{fig:splitdensities-3L}\includegraphics[width=8.2cm]{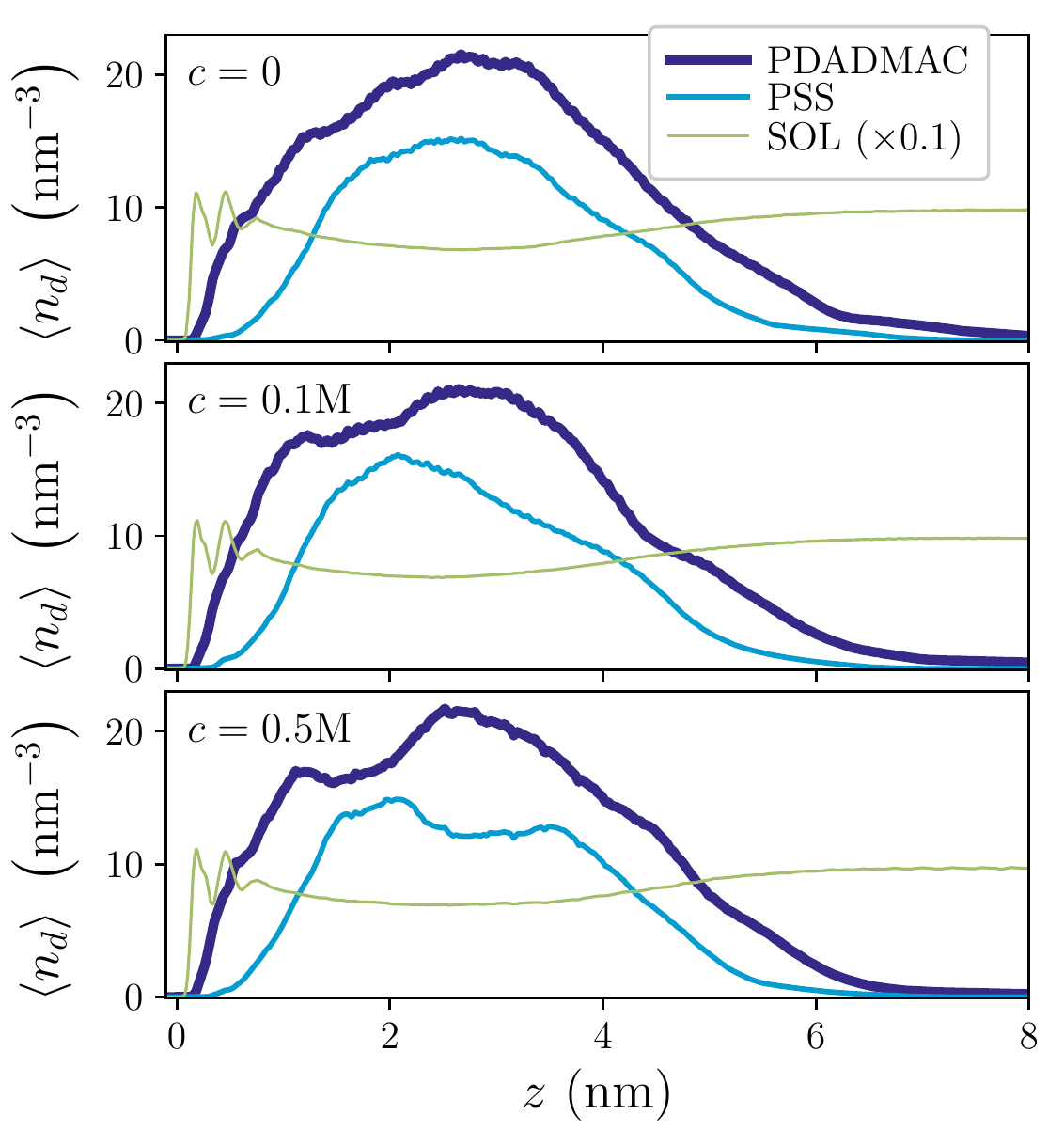}}
\subfigure[]{\label{fig:splitdensities-4L}\includegraphics[width=8.2cm]{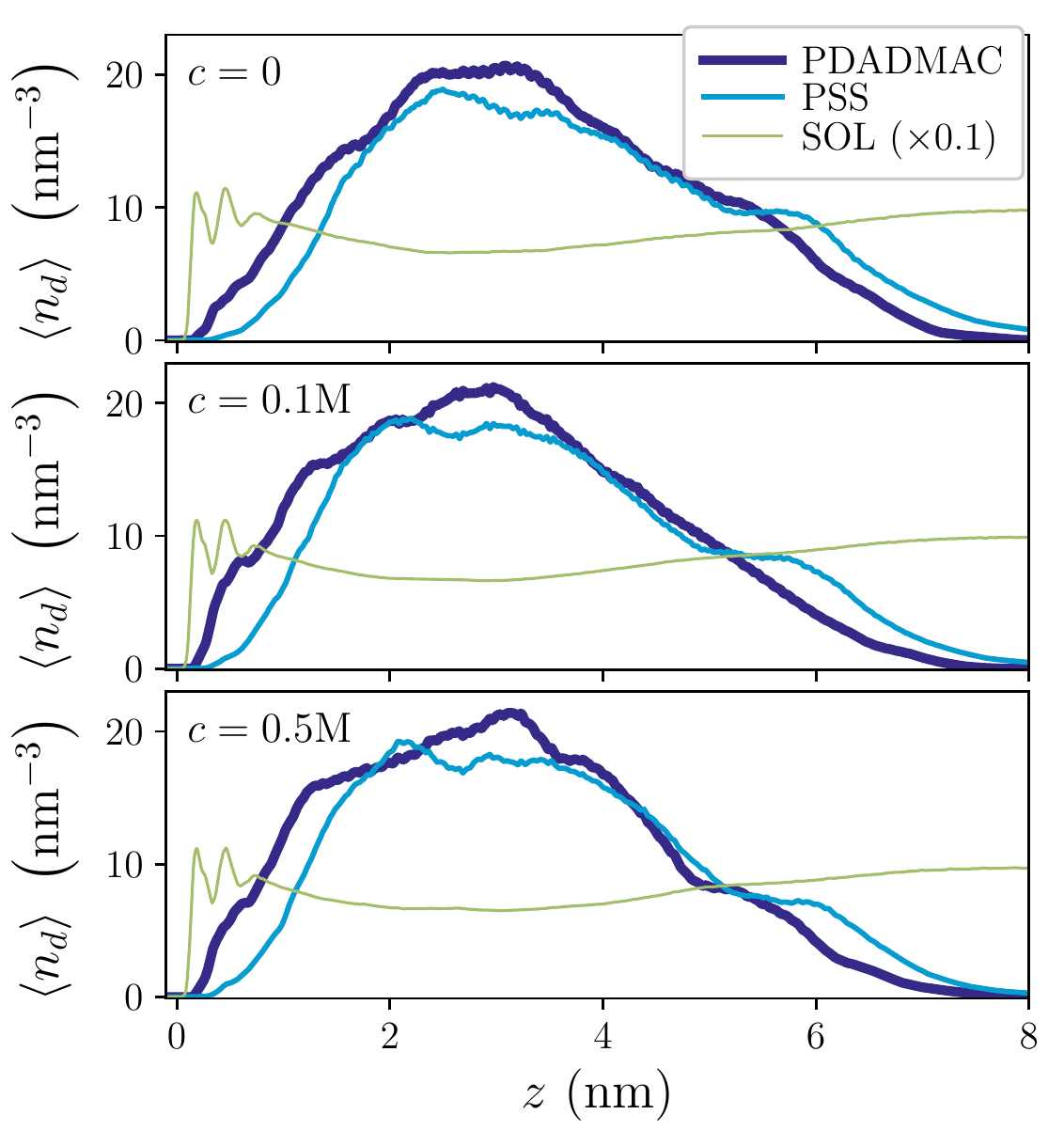}}
\caption{Average split number density profiles corresponding to each OE type and water. (a) Results for 3L. (b) Results for 4L.}
\label{fig:splitdensities}
\end{figure}

In order to have a more detailed view of the structure of the OEMs, we analyze the density distributions of atoms belonging to PDADMAC or PSS chains separately. Such OE split number density profiles are shown in Figures~\ref{fig:splitdensities-3L} and \ref{fig:splitdensities-4L}, together with the water profiles introduced above as a reference.
All these profiles evidence that, as expected, PDADMAC tends to be closer to the substrate surface than PSS, whereas in the outermost region there is a higher presence of PDADMAC in 3L systems and of PSS in 4L ones. Also as expected, the 3L profiles indicate a significantly higher presence of atoms belonging to PDADMAC than that corresponding to PSS within the whole structure of the film (Figure~\ref{fig:splitdensities-3L}). In 4L, instead, profiles of both OEs are quantitatively very similar. However, when comparing number density profiles of each OE from the same system one has to keep in mind that PDADMAC chains have a larger number of atoms than PSS ones with the same length (with the selected degree of polymerization, 752 and 572, respectively). In fact, the average total amount of PDADMAC chains forming the films is $17\pm2$, in front of the $12\pm 1$ of PSS chains in 3L systems and $21\pm 2$ in 4L ones. The difference in atoms content of each OE type also prevents us from distinguishing a potentially alternating layer structure for them. This aspect will be analyzed below by means of more detailed parameters.

\subsubsection{OEs layering: center-of-mass distributions}
\begin{figure}[!t]
\centering
\subfigure[]{\includegraphics[width=8.2cm]{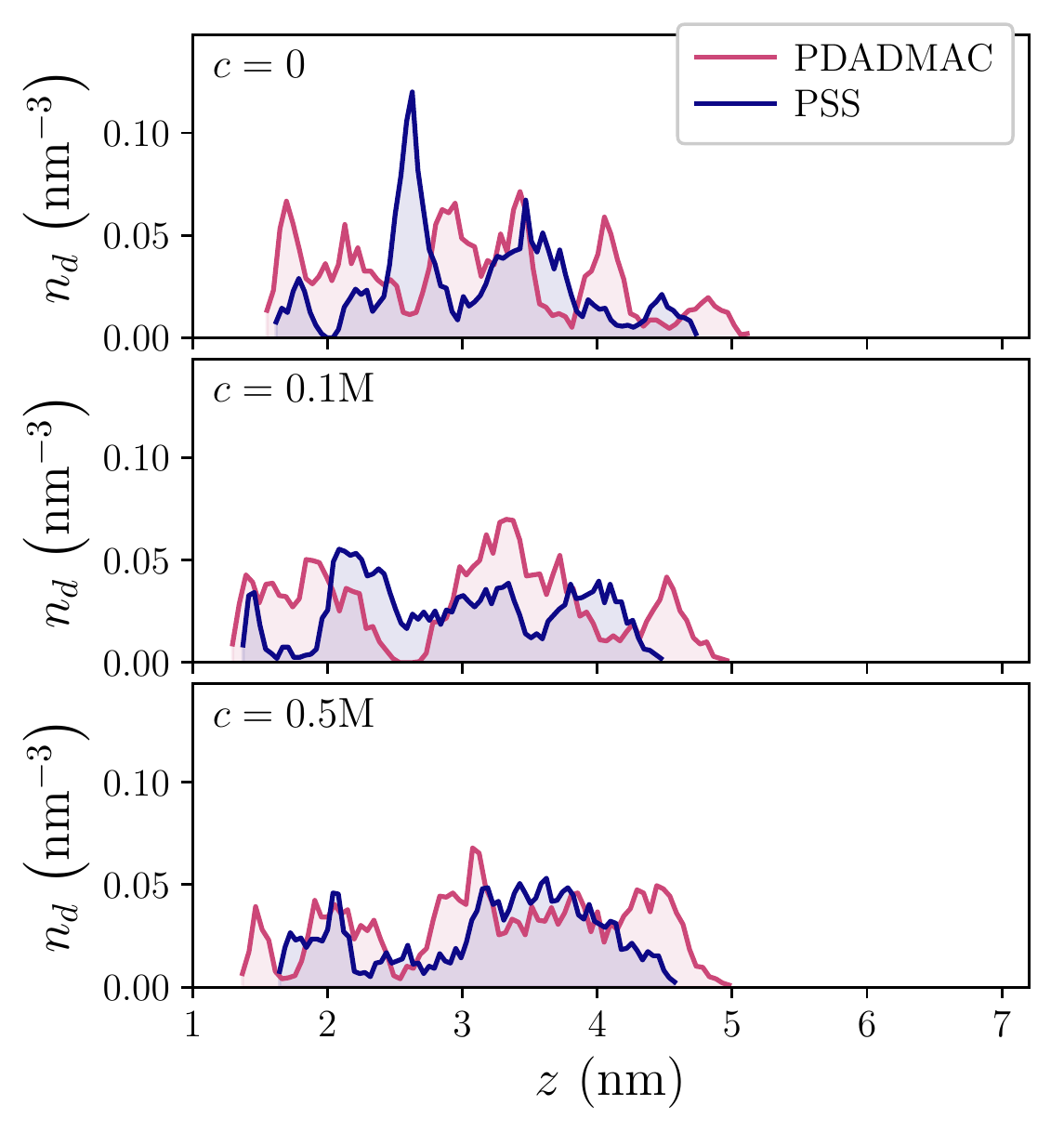}}
\subfigure[]{\includegraphics[width=8.2cm]{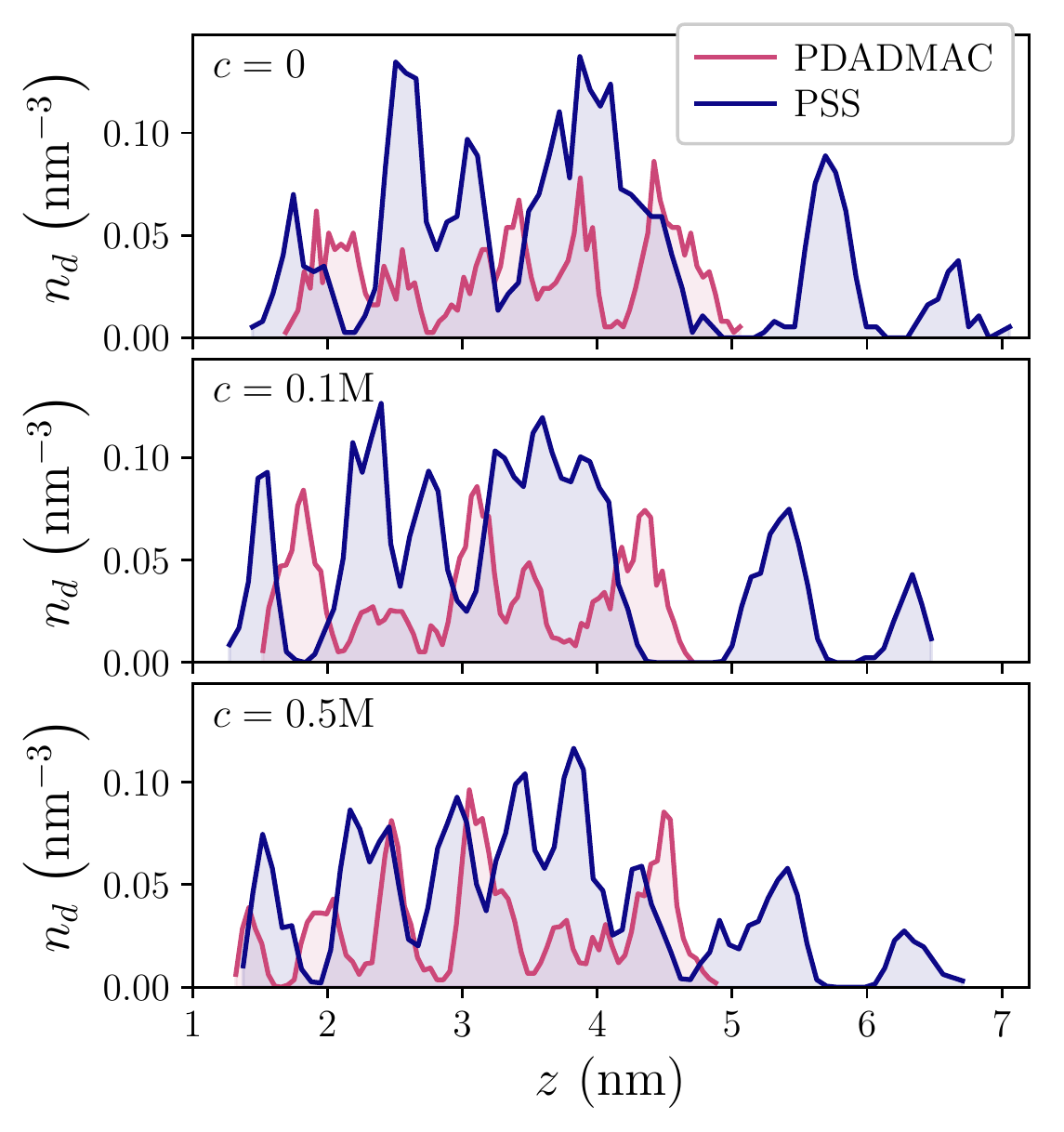}}
\caption{Examples of density profiles of OEs center of mass obtained in selected simulation runs for systems 3L (a) and 4L (b) for each added salt concentration.}
\label{fig:densities-layers}
\end{figure}
In order to clarify whether our simulated OEMs have an alternating layered internal structure, we calculated the split density profiles of the centers of mass of individual OE molecules. In this way, we avoid the aforementioned issue related to the different atom content of the pair of OEs used here. Figure~\ref{fig:densities-layers} shows some examples of the results obtained in such calculation from selected individual simulation runs. Alternating maxima are clearly visible in all cases, even though the profiles are rather noisy due to the relatively low amount of OE chains involved. Interestingly, the amount of alternating maxima is not equal to the corresponding amount of deposition cycles from which each sample was obtained, but slightly higher. This indicates a rather entangled arrangement of oppositely charged OE chains, as the snapshots in Figure~\ref{fig:snapshots} also suggest. This is an unexpected result, since the short length of these OE chains can not favor strong entanglements in the system. Additionaly, this lack of order in the internal structure of OEMs contradicts, in principle, the hypothesis of a structure of well defined flat layers that was made in reference \cite{2014-micciulla-pccp} for these systems. Such hypothesis was based on the experimental measurements of the mechanical properties of the films and the structural model proposed in reference \cite{2013-nestler}.

An explanation for the high entanglement observed in our simulated samples is the very low substrate surface coverage obtained during the first adsorption cycles (see Figure~\ref{fig:snapshots}). As was already discussed in reference \cite{2014-micciulla-softmat}, the coverage fraction grows from approximately 30\% of the surface area after the first deposition cycle to nearly 80\% after the fourth. Thus, the OE chains adsorbed during the second and, to a decreasing extent, the third and fourth cycles reach the inner regions of the film, leading to a rearrangement of the firstly deposited layers and the formation of complex internal configurations. For higher deposition cycles, one can expect the innermost region to become unreachable for newcoming chains due to the crowded central region. The slower decay of the density profiles in the outermost region of 3L samples compared to 4L ones, as seen in Figure~\ref{fig:ovalldensprof}, suggests that the free surface of the film may become flatter with larger amounts of deposition cycles, reducing the entanglements, thus favoring the layered structure. Therefore, here we only have indications on the hypothesis of a high alternating layering in OEMs internal structure not being valid for the first few deposition cycles, whereas it might still hold for layers above the few ones closest to the substrate. This would imply that the structure of OEMs near the substrate can be rather different than that of outer regions. As pointed above, such distinct structure near the substrate is determined by the strong hydration of the surface of the latter. Finally, an alternative or complementary explanation to the observed lower entanglement than the one predicted from experimental measurements could be the ordering effect of eventual slow OE diffusion processes within the film\cite{2017-fares}. Unfortunately, such slow processes are far beyond the time scales that are reachable in our simulations.

\subsubsection{OE chain configurations}
\begin{figure}[!t]
\centering
\subfigure[]{\label{fig:Rg-DPs}\includegraphics[width=8.2cm]{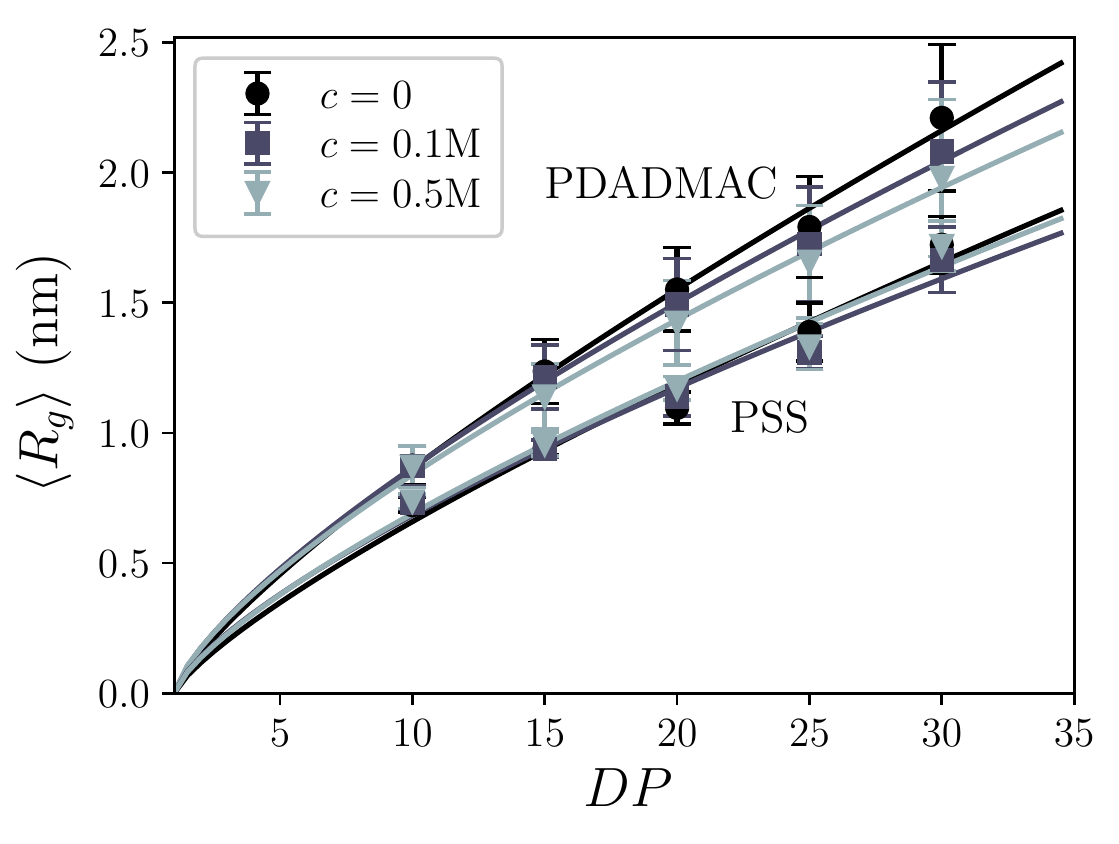}}
\subfigure[]{\label{fig:Rg-DP30}\includegraphics[width=8.2cm]{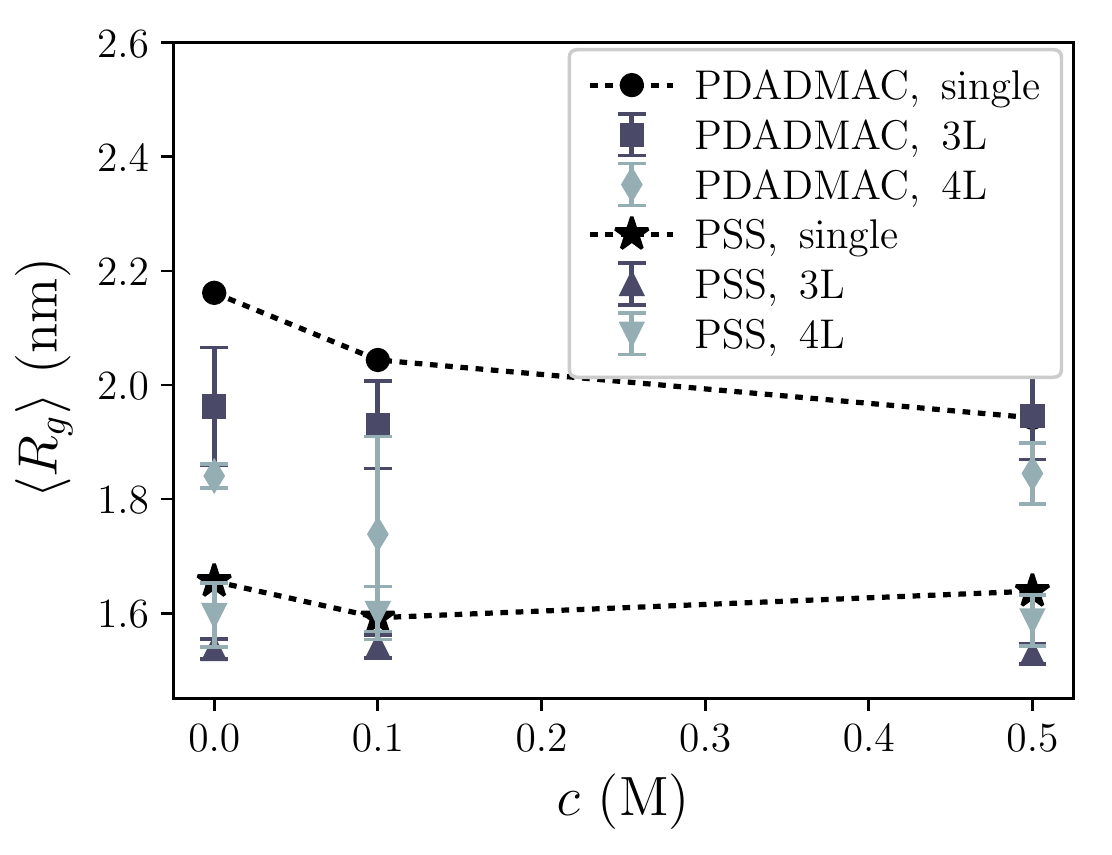}}
\caption{Average radius of gyration, $\langle R_g \rangle$, of OE chains. (a) Results from simulations of single OE chains with different degrees of polymerization (points) and corresponding least squares fits to a power law function (solid lines). Fitted exponents are, for increasing values of $c$, $\alpha=\lbrace 0.78,\, 0.73,\, 0.72\rbrace$ for PDADMAC and $\alpha=\lbrace 0.79,\, 0.72,\, 0.74\rbrace$ for PSS, respectively. (b) Comparison of the values of $\langle R_g \rangle$ for single chains with $DP=30$, obtained from the fitted functions, and for chains in 3L and 4L OEMs.}
\label{fig:Rgs}
\end{figure}

After the examination of the overall structure and the arrangement of polyanion and polycation chains within the multilayer, we compare the average structure of such chains to the reference case corresponding to polyanions and polycations at infinite dilution. Even though the structural properties of high molecular weight PDADMAC and PSS chains are well known,\cite{2009-adamczyk-csa, 2014-adamczyk-jcis, 2005-marcelo-pol} these cannot be extrapolated to such low molecular weight oligomers as the ones sampled here. Since oligomers are not expected to follow the same scaling laws, we need to perform a direct characterization of their structure. In order to do this, first we calculate the radius of gyration of the chains, $R_g$. It is defined as
\begin{equation}
 R_g = \left ( \lambda_1^2+ \lambda_2^2 + \lambda_3^2 \right )^{1/2},
\end{equation}
where $\lambda_i$ are the eigenvalues of the gyration tensor, $\tilde Q$, corresponding to the positions of all atoms belonging to a single chain. The elements of $\tilde Q$ are defined as $Q_{mn} = \frac{1}{N} \sum_{i=1}^N r_m^{(i)} r_n^{(i)}$, where $r_m^{(i)}$ is the $m$-th cartesian coordinate of the $i$-th atom in the chain and $N$ is the total amount of atoms. The results of $R_g$ averaged over OE chains, sampled configurations and independent runs, $\langle R_g \rangle$, are plotted in Figure~\ref{fig:Rgs}.

Specifically, Figure~\ref{fig:Rg-DPs} corresponds to the values of $\langle R_g \rangle$ obtained from simulations of single PDADMAC and PSS chains with degrees of polymerization ranging from 10 to 30 under the three sampled added salt concentrations. Least squares fits to the power law function $R_g=k (DP-1)^{\alpha}$ for each value of $c$ are also included. Fitted exponents obtained for both polymers are around $\alpha \sim 0.75$. As expected, these values differ significantly from the behavior of large $DP$ systems. For instance, it is known that $\alpha \approx 0.51$ for high molecular weight PDADMAC under 0.5M of NaCl.\cite{2005-marcelo-pol} Figure~\ref{fig:Rg-DPs} also shows that for any sampled $DP$, PDADMAC chains have a larger extension than PSS ones in all cases. For PDADMAC, one can see that the extension also grows slightly with decreasing $c$. This is an expected behavior, as it can be attributed to the screening of the repulsion between charged groups along the chain backbone by ion condensation, that is favored by higher added salt concentrations. However, for PSS chains the effect of $c$ is too small to be clearly distinguished.

In Figure~\ref{fig:Rg-DP30} we compare the values of $\langle R_g \rangle$, corresponding to $DP=30$, obtained from the simulations of single OE chains and the simulations of 3L and 4L OEMs. In this case it can be observed that chains within OEMs tend to be slightly less extended than the ones under infinite dilution conditions. This difference is clearer for PDADMAC chains, but in any case the change in $\langle R_g \rangle$ is not bigger than approximately 10\%. Even the dependence of the observed differences on the amount of deposition cycles compares to the width of the error bars, it seems that there is a qualitative alternating effect on the average values: PDADMAC chains tend to be less extended in 4L than in 3L systems, whereas PSS chains show the opposite trend. This suggests a slight dependence of the charge compensation mechanisms of each type of OE on the amount of deposition cycles. However, for such short chain lengths, the characteristic extension of the OEs does not seem to be strongly affected by being part of the thin film structure.

Besides the relative extension of the chains, one expects their orientation to be significantly affected by the geometrical constraints of the multilayer structure. In a highly layered structure, one would expect the OE chains to adopt rather flat configurations mainly parallel to the plane defined by the substrate surface. In order to decide this question, we analyze two orientational parameters. The first one is a measure of the degree of parallelism of the OE chain with respect to the plane of the substrate surface. We estimate such degree as the angle between the substrate surface and the major rotation axis of the chain,
\begin{equation}
\theta=\frac{\pi}{2} - \arccos \left ( \frac{\vec a_1 \cdot \hat n} {\| \vec a_1 \|} \right )
\end{equation}
where $\hat n$ is the unitary vector perpendicular to the substrate plane and $\vec a_1$ is the major eigenvector of the inertia tensor of the polymer, $\tilde I$. The latter is defined as $\tilde I = \sum_{i=1}^N m_i \left [ (\vec r_i \cdot \vec r_i ) \vec I - \vec r_i \otimes \vec r_i \right ]$, where $m_i$ and $\vec r_i$ are the mass and position of the $i$-th atom and $\vec I$ is the identity matrix. Therefore, this parameter will take the value $\theta=0$ when the main rotation axis of the chain is perfectly parallel to the substrate surface and $\theta=\pi/2$ when it is perpendicular. The second parameter is the prolateness of the chain, $S$, which characterizes the shape asymmetry of its conformation, assuming it fits a spheroidal volume distribution. $S$ is also defined from the eigenvalues of the gyration tensor as
\begin{equation}
 S = \frac {\prod_{i=1}^3 \left ( \lambda_i - \langle \lambda_i \rangle \right )}{\langle \lambda_i \rangle^3}.
\label{eq:prolateness}
\end{equation}
This parameter takes values between -1 and 1, classifying three types of shapes: for $S>0$ the shape is prolate, which corresponds to a spheroid generated by the rotation of an ellipse around its major axis; for $S=0$ the shape is spherical; for $S<0$ the shape is oblate, corresponding to a spheroid obtained from the rotation of an ellipsoid around its shorter axis.
\begin{figure}[!t]
\centering
\includegraphics[width=8.2cm]{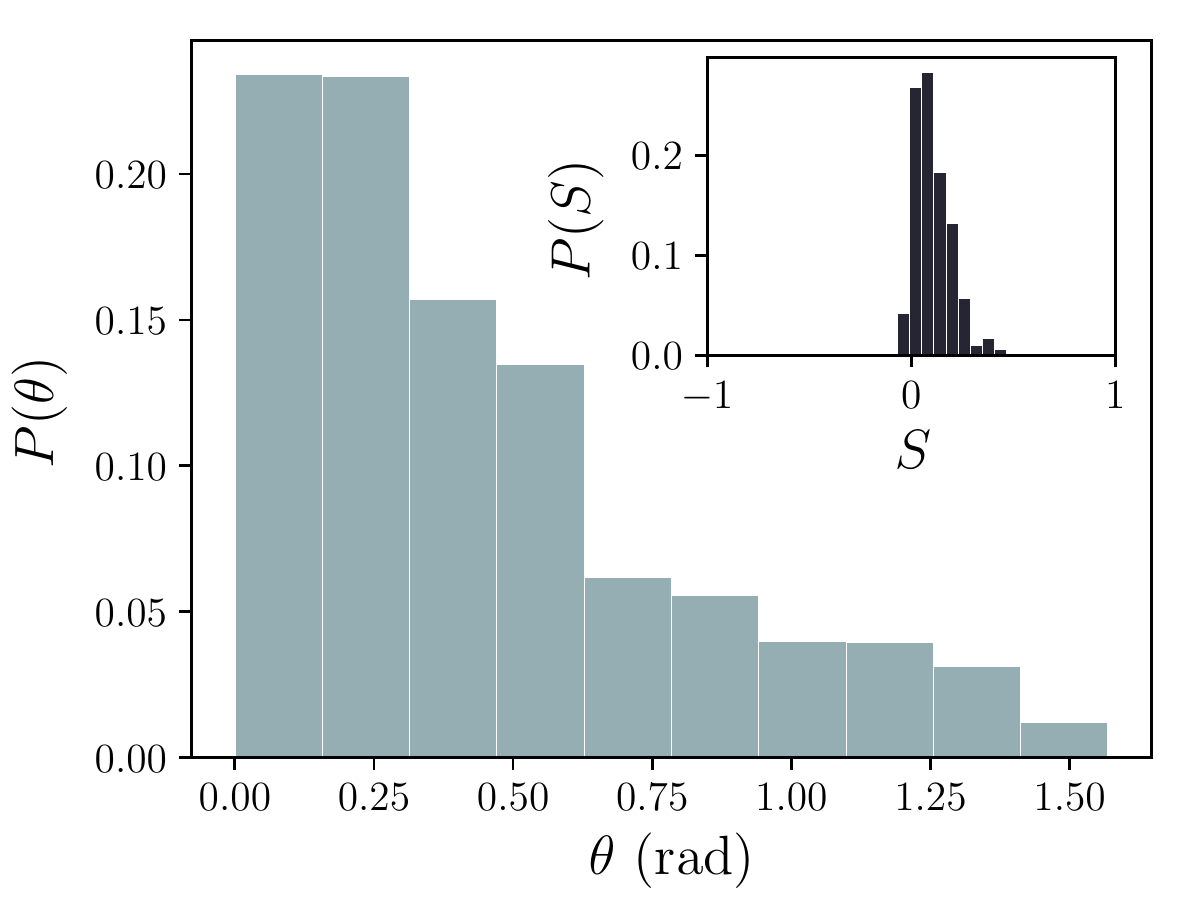}
\caption{Probability distributions for the degree of parallelism with respect to the plane of the substrate (main plot) and for the prolateness (inset) of OE chains in 4L systems under 0.5M of added salt conditions.}
\label{fig:shapeparams}
\end{figure}
Figure~\ref{fig:shapeparams} shows the distributions of probability $P(\theta)$ and $P(S)$ of such parameters for the case of OE chains belonging to 4L systems exposed to a concentration of added salt of 0.5M. The maximum probability for $\theta$ clearly corresponds to $\theta \sim 0$, and strongly decays for larger angles. This confirms that, despite the entangled structure of the multilayer, polymers tend to orient their major axis of inertia parallel to the substrate surface. Regarding $S$, its distribution shows that the chains only adopt slightly prolate shapes. Similar distributions (not shown) have been obtained for the rest of sampled systems.

In conclusion, the conformations of individual OE chains within the films are slightly more compact than in the case of infinite dilution conditions, keeping a rather parallel orientation with respect to the substrate surface despite the fact that oppositely charged chains tend to entangle.

\subsection{Charge properties}
The structural properties discussed above are mainly determined by electrostatic interactions in the system. Here we analyze the charge distributions, charge compensation mechanisms, overcharge fractions and growth regime corresponding to our simulated multilayers.

\subsubsection{Split and net charge distributions}
\begin{figure}[!t]
\centering
\subfigure[]{\label{fig:chargedists-3L}\includegraphics[width=8.2cm]{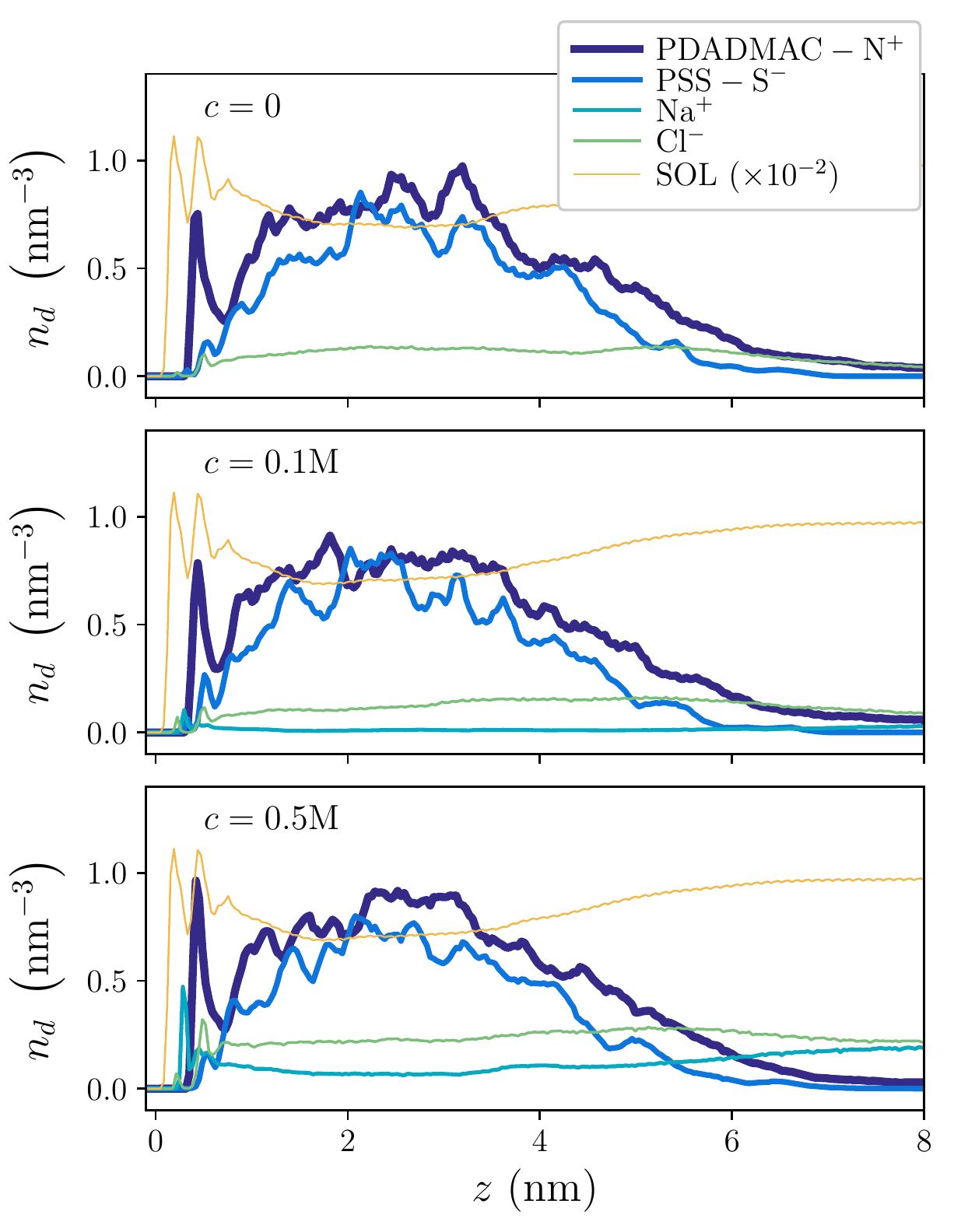}}
\subfigure[]{\label{fig:netcharge-3L}\includegraphics[width=8.2cm]{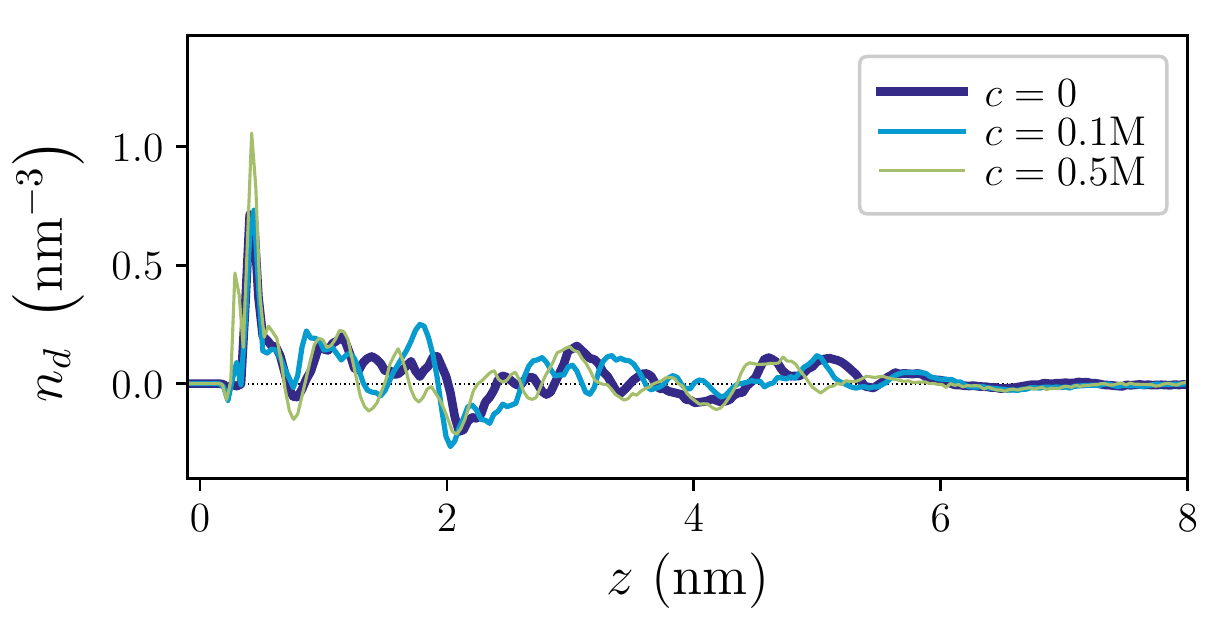}}
\caption{Number density profiles of charges and water in selected 3L systems for each sampled added salt concentration. (a) Split charge distributions. (b) Net charge distributions, excluding substrate surface charges.}
\label{fig:chargedistsnet-3L}
\end{figure}

\begin{figure}[!t]
\centering
\subfigure[]{\label{fig:chargedists-4L}\includegraphics[width=8.2cm]{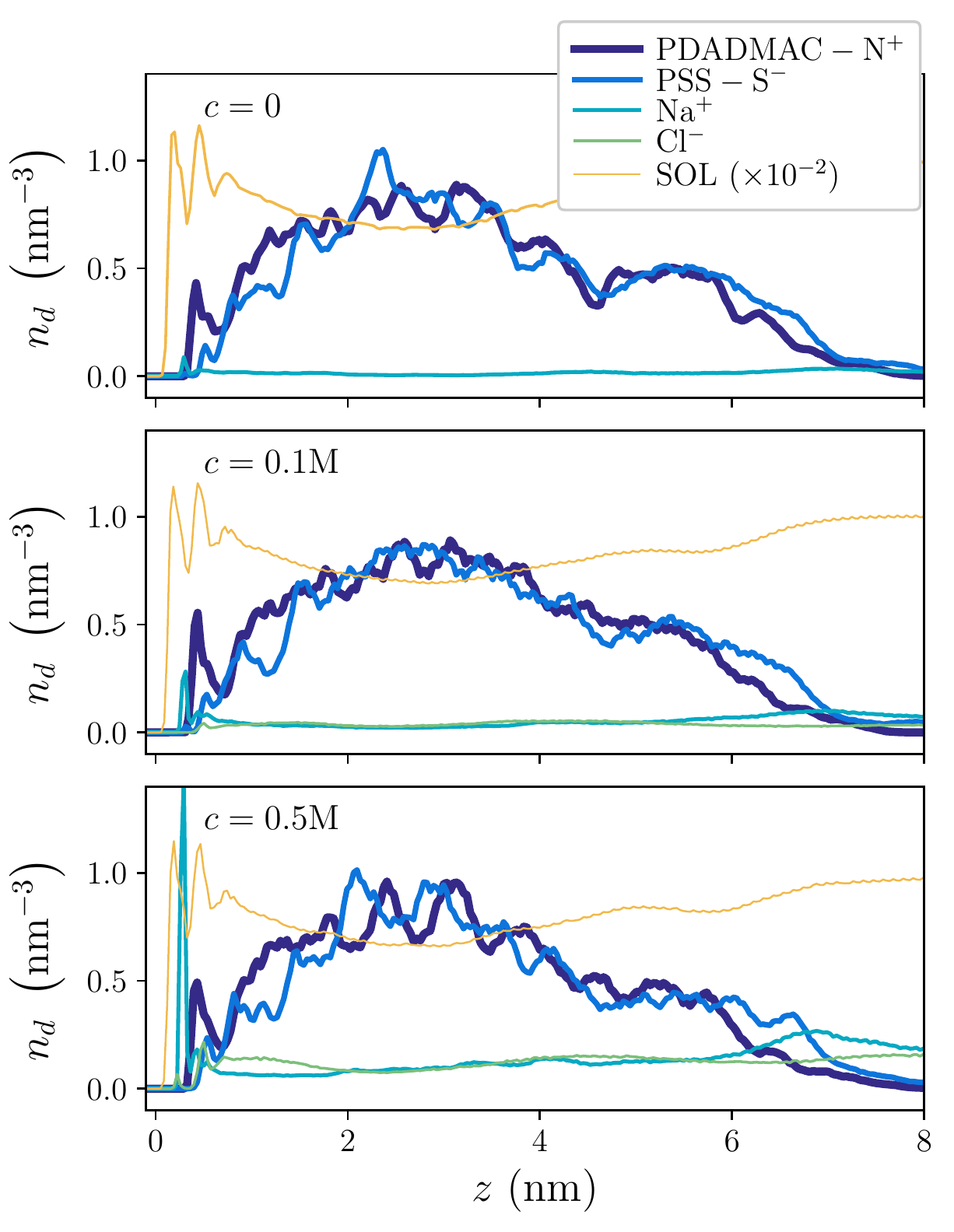}}
\subfigure[]{\label{fig:netcharge-4L}\includegraphics[width=8.2cm]{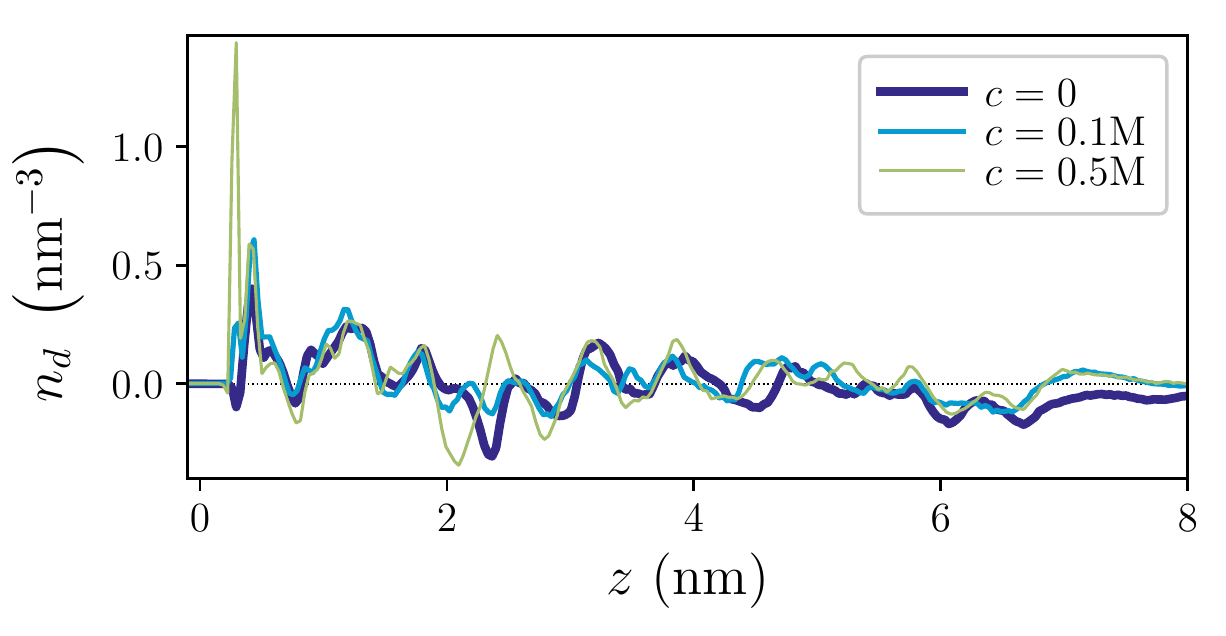}}
\caption{Number density profiles of charges and water for 4L selected systems. (a) Split charge distributions. (b) Net charge distributions, excluding substrate surface charges.}
\label{fig:chargedistsnet-4L}
\end{figure}

First, we analyze the distributions of charges as a function of the distance to the substrate. Figures~\ref{fig:chargedistsnet-3L} and \ref{fig:chargedistsnet-4L} include different charge density profiles corresponding to selected, not averaged, 3L and 4L systems, respectively. Specifically, Figures~\ref{fig:chargedists-3L} and \ref{fig:chargedists-4L} show the split number density profiles of each charged component in the system, \textit{i.e.}, Na$^+$ and Cl$^-$ salt ions and the nitrogen and sulfur atoms from PDADMAC and PSS OEs, whereas Figures~\ref{fig:netcharge-3L} and \ref{fig:netcharge-4L} correspond to the distributions of net charge in each system. In the former plots, the density profiles of water have been also included for the convenience of the following discussion. One has also to note that, for $c=0$, sodium cations are absent because the equilibration in pure water adds no ions to the system after the rinsing process, which removes all ions that do not serve as counter-ions of the excess of OE charged groups. According to the LbL growth mechanism, the latter necessarily correspond to a part of the OEs adsorbed during the last deposition cycle.

As we discussed for the split density profiles above, one can distinguish three different regions for the charge profiles in Figures~\ref{fig:chargedists-3L} and \ref{fig:chargedists-4L}: a narrow, highly layered region close to the substrate, a broad central region with high density of OEs and a region of slowly decaying density corresponding to the film free surface. Starting with the profiles of the 3L system (Figure~\ref{fig:chargedists-3L}), we can see that charges close to the substrate show a strong layering in all cases, as it happens with water. PDADMAC charged groups exhibit a prominent single peak of high density that grows slightly with $c$. This corresponds to the electrostatic attraction of the -N$^+$ groups to the negatively charged hydroxyl groups that sightly emerge from the substrate surface. The position of such peak approximately coincides with the second main peak in water profile. Above this peak, the profile of -N$^+$ groups shows a gap of low density that extends up to the beginning of the central region of the film. This suggests that only few PDADMAC chains connect the dense narrow layer close to the substrate with the dense central region. However, PSS charged groups in this innermost region show a rather different profile: due to the electrostatic repulsion they experience from the substrate surface, their density is very low and only grows slowly, with moderate fluctuations, as the distance to the substrate increases. The same electrostatic argument explains the higher presence next to the substrate of $\mathrm{Na}^+$ ions in front of $\mathrm{Cl}^-$ ones when the added salt concentration is not very low: at $c=0.5$M one can see that ions also exhibit a significant layering in their profiles, with two asymmetric peaks. For $\mathrm{Cl}^-$, positions of these peaks are very close to the ones corresponding to water layers, whereas for $\mathrm{Na}^+$ the most prominent peak is shifted to a point in between the latter, slightly below the layer of PDADMAC charges. This corresponds to a partial but significant ion condensation on the charged groups of the substrate and the adsorbed monomers of PDADMAC chains, that strongly depends on the concentration of added salt. The central regions of 3L films show a much less structured profile, with a rather flat distribution of salt ions and a high density of OE charged groups. The latter exhibit moderate oscillations whose length is similar to the one observed for the peaks in the distribution of centers of mass shown in Figure~\ref{fig:densities-layers}. However, the slow decay to zero of the density of OE charges in the outermost region of the films is relatively smooth, Regarding 4L systems, the main difference with respect to 3L ones in their split charge density profiles lies only in the region close to the substrate. In this case the single peak of PDADMAC charged groups is much less pronounced and shows no clear dependence on the added salt concentration. However, the main peak of Na$^+$ ions in 4L systems is significantly larger and shows a much stronger growth with $c$. This means that the electrostatic attraction from the substrate surface on the PDADMAC charged groups experiences a stronger screening from sodium cations in PSS ended films. Finally, the net charge distributions plotted in Figures~\ref{fig:netcharge-3L} and \ref{fig:netcharge-4L} show multiple alternating positive and negative small peaks along the film, except in the region close to the substrate. Taking into account that, with the same exception, the distributions of ions are rather flat, such multiple alternating maxima are due to the fluctuations of OEs charged groups. As in the profiles of OEs centers of mass (Figure~\ref{fig:densities-layers}), this reflects the rather entangled structure of OE chains in the film. Next to the substrate, the net charge profiles follow what has been pointed out for the split distributions: a strong layering, very sensitive to added salt concentration and type of film ending.

\subsubsection{Charge pairing}
\begin{figure}[!t]
\centering
\subfigure[]{\label{fig:chargecomp-intrinsic}\includegraphics*[width=8.4cm]{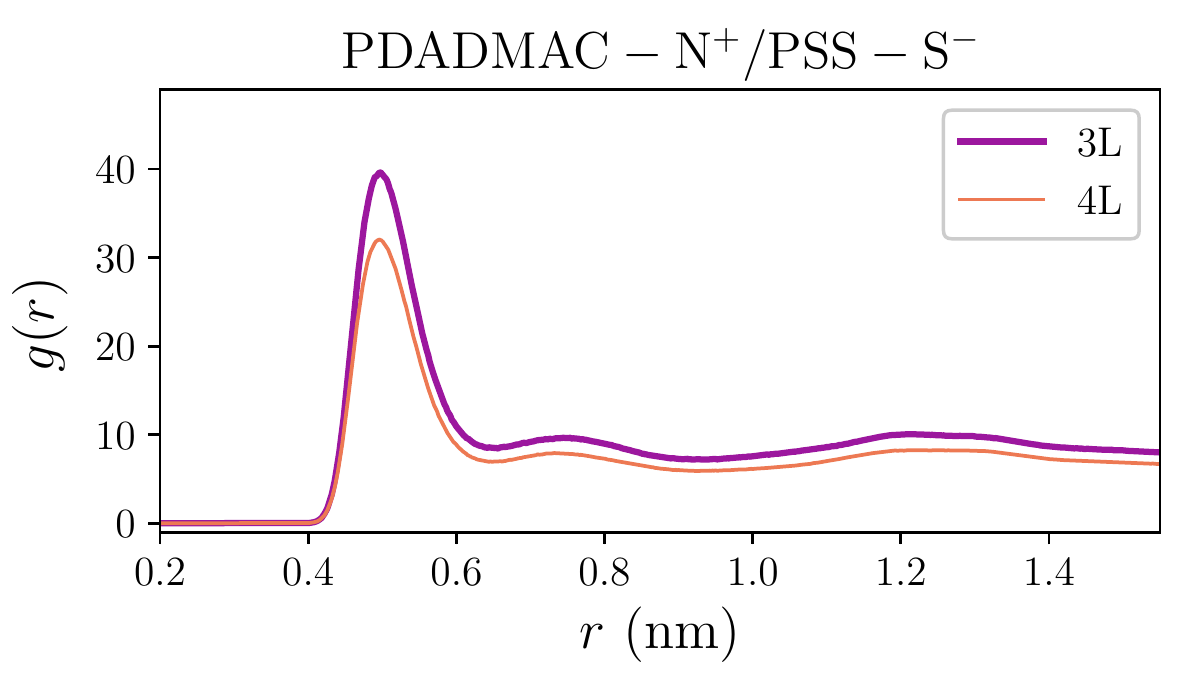}}
\subfigure[]{\label{fig:chargecomp-extrinsic}\includegraphics*[width=8.4cm]{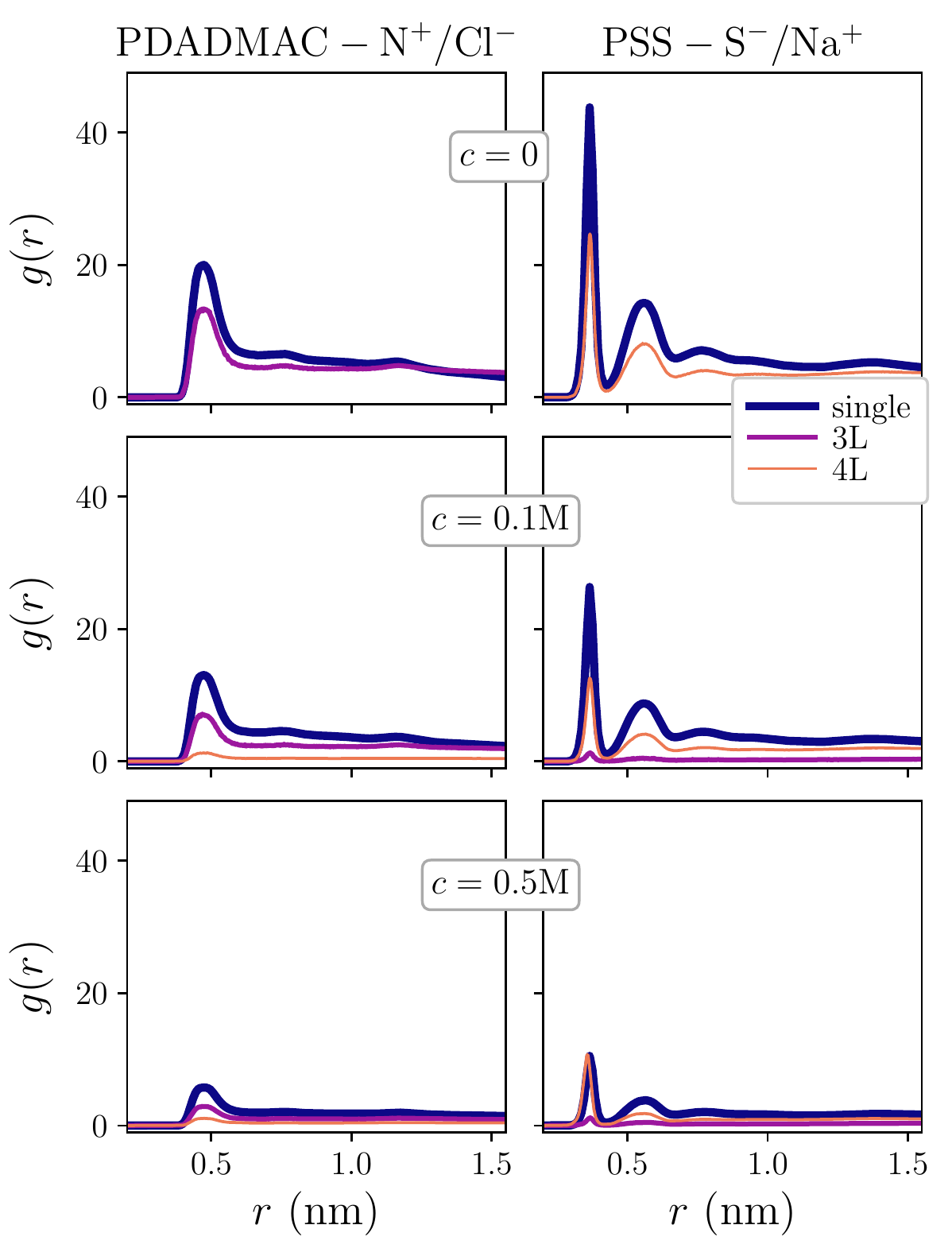}}
\caption{Pair correlation functions for intrinsic (a) and extrinsic (b) charge compensation. Only the case $c=0$ is shown in (a), as curves for $c=0.1$M and $c=0.5$M are very similar.}
\label{fig:chargecomp}
\end{figure}

PE charged groups in assembled complexes and multilayers can compensate their charges by means of two mechanisms: the `intrinsic compensation', that consists in the pairing of oppositely charged PE groups, and the `extrinsic compensation', that is the pairing of such groups with any free ion present in the system, either counter-ions or added salt ions\cite{2002-riegler}. In simulation data, the importance of each charge compensation mechanism can be determined from ion pairing correlation functions. Figures~\ref{fig:chargecomp-intrinsic} and \ref{fig:chargecomp-extrinsic} show the radial distribution functions corresponding to extrinsic and intrinsic charge compensation, respectively, in both 3L and 4L OEMs. Curves corresponding to extrinsic charge compensation of OE chains at infinite dilution are also included in Figure~\ref{fig:chargecomp-extrinsic}. A direct comparison of the heights of the first peak in each distribution reveals that the main charge compensation mechanism in all OEMs systems is intrinsic. This mechanism turns out to be rather insensitive to $c$, as curves corresponding to $c=0.1$ and 0.5M (not shown) are very similar to the ones in Figure~\ref{fig:chargecomp-intrinsic}, that correspond to $c=0$. As mentioned above, the short times reached in atomistic simulations can not capture the very slow dynamics of salt annealing processes involving significant rearrangements of PE chains, whose typical time scales are measured in hours\cite{2001-dubas}. Therefore, in our results, intrinsic compensation is mainly affected by the amount of deposited layers: as we can see in Figure~\ref{fig:chargecomp-intrinsic}, the distribution function is slightly lower for 4L than for 3L. This suggests the existence of a higher overcharge in the outermost region of 4L systems, that implies a relatively higher amount of PSS-S$^-$ groups not intrinsically compensated. On the other hand, the dependence of extrinsic compensation of each type of PE charged groups on the deposited layers clearly follows the expected behavior: as Figure~\ref{fig:chargecomp-extrinsic} shows, the largest extrinsic compensation corresponds to PEs forming the outermost layer in each system (PDADMAC-N$^+$ for 3L, PSS-S$^-$ for 4L). Obviously, extrinsic compensation is weaker in OEMs than in single PE systems due to the effect of intrinsic compensation, that is absent in the latter. Finally, one can see a decrease of the heights of the curves as $c$ increases, in correspondence to an increase of Na$^+$-Cl$^-$ ion pairing in 3L (not shown) and to an increasingly large accumulation of Na$^+$ ions next to the substrate in 4L (Figure~\ref{fig:chargedistsnet-4L}).

\begin{table}[!t]
\small
\centering
\caption{\label{tab:firstcoord}First coordination number, $n_1$, for different ion pairings ($\pm 0.1$, standard deviation), where -N$^+$ denotes PDADMAC-N$^+$ and -S$^-$ refers to PSS-S$^-$.}
%\begin{tabular}{c c c c c c}
\begin{tabular*}{0.48\textwidth}{@{\extracolsep{\fill}}c c c c c c}
   \hline
   system & $c$ (M)   & Na$^+$/Cl$^-$    & -N$^+$/Cl$^-$ & -S$^-$/Na$^+$  & -N$^+$/-S$^-$ \\
   \hline
   \begin{tabular}[c]{@{}c@{}}~\\ single\\ ~\end{tabular} & \begin{tabular}[c]{@{}c@{}}0\\ 0.1\\ 0.5\end{tabular} & \begin{tabular}[c]{@{}c@{}}-\\ -\\ -\end{tabular} & \begin{tabular}[c]{@{}c@{}}3.0\\ 1.9\\ 0.8\end{tabular} & \begin{tabular}[c]{@{}c@{}}1.7\\ 1.0\\ 0.4\end{tabular} & \begin{tabular}[c]{@{}c@{}}-\\ -\\ -\end{tabular}\\
   \hline
   \begin{tabular}[c]{@{}c@{}}~\\ 3L\\ ~\end{tabular} & \begin{tabular}[c]{@{}c@{}}0\\ 0.1\\ 0.5\end{tabular} & \begin{tabular}[c]{@{}c@{}}-\\ 0.1\\ 0.2\end{tabular} & \begin{tabular}[c]{@{}c@{}}2.8\\ 1.4\\ 0.6\end{tabular} & \begin{tabular}[c]{@{}c@{}}-\\ 0.1\\ 0.1\end{tabular} & \begin{tabular}[c]{@{}c@{}}10.0\\ 10.7\\ 10.2\end{tabular}\\
   \hline
   \begin{tabular}[c]{@{}c@{}}~\\ 4L\\ ~\end{tabular} & \begin{tabular}[c]{@{}c@{}}0\\ 0.1\\ 0.5\end{tabular} & \begin{tabular}[c]{@{}c@{}}-\\ 0.1\\ 0.1\end{tabular} & \begin{tabular}[c]{@{}c@{}}-\\ 0.2\\ 0.2\end{tabular} & \begin{tabular}[c]{@{}c@{}}0.7\\ 0.3\\ 0.3\end{tabular} & \begin{tabular}[c]{@{}c@{}}8.8\\ 8.9\\ 9.0\end{tabular}\\
   \hline
\end{tabular*}
\end{table}
A quantitative analysis can also be performed on charge pairing distributions by calculating the first coordination number, $n_1$, of each pair. In general, we can define the coordination number of the $i$-th coordination shell as
\begin{equation}
 n_i = 4\pi \rho \int_{r_{\mathrm{min}_i}}^{r_{\mathrm{min}_{i+1}}} r^2 g(r) dr,
 \label{eq:coordnum}
\end{equation}
where $\rho$ is the number density of the involved species within the film region. We evaluate expression~\ref{eq:coordnum} by numerical integration of the area under $g(r)$ comprised between the positions of the two minima bounding the $i$-th peak, $r_{\mathrm{min}_{i}}$ and $r_{\mathrm{min}_{i+1}}$. That is, $n_1$ is given by the area under $g(r)$ from the position of its first non-zero value up to the position of the minimum after the first peak. The values obtained from this calculation for each relevant case are included in Table~\ref{tab:firstcoord}. We can see that the values corresponding to the pair PDADMAC-N$^+$/PSS-S$^-$ are significantly larger than for any other pair, confirming the dominance of the intrinsic charge compensation in all OEMs. These values also confirm that such intrinsic compensation is slightly lower for 4L systems, in contradiction with the general higher intrinsic compensation of PSS ended PEMs and OEMs postulated indirectly from experimental observations\cite{2014-micciulla-pccp}. However, this contradiction might be partially alleviated when one observes that PSS-S$^-$ has also a lower extrinsic compensation in 4L than PDADMAC-N$^+$ in 3L, which is a behavior specifically observed experimentally in analogous PEM systems\cite{2013-ghostine}. Thus, we can only conclude that, in average, our 4L systems have been slightly less efficient in charge pairing than 3L ones. In any case, one has to keep in mind that our results still correspond to a transient growth regime and that the small details might be dominated by fluctuations.

\subsubsection{Overcharge fraction and growth regime}
Experimentally, two types of PEM growth can be observed: linear growth, in which the amount of adsorbed PEs in each deposition cycle is independent from the total number of deposited layers, and exponential growth, in which such amount increases with the number of layers. A common way to determine the growth regime achieved in simulations is by means of the overcharge fraction\cite{2006-patel}, that is defined as
\begin{equation}
 f_n = \left | \frac{\sum_{i=0}^n Q_i}{Q_n} \right |,
 \label{eq:overcharge}
\end{equation}
where $Q_n$ is the charge of the PE chains adsorbed in the deposition cycle $n$ and $Q_0$ is the charge of the substrate. It is assumed that a linear growth is obtained for $f_n \approx 0.5$, an exponential growth for $f_n > 0.5$ and no stable growth occurs for $f_n < 0.5$.

\begin{table}[!b]
\small
\centering
\caption{\label{tab:overcharge}Overcharge fraction for every deposited layer. Intervals correspond to standard deviations.}
\begin{tabular*}{0.48\textwidth}{@{\extracolsep{\fill}}c c c c}
   \hline
   & $n$L   &  $f_n$ & \\
   \hline
   & 1L   &  0.80 $\pm$0.05 & \\
   & 2L   &  0.69 $\pm$0.04 & \\
   & 3L   &  0.25 $\pm$0.05 & \\
   & 4L   &  0.8 $\pm$0.1  & \\
   \hline
\end{tabular*}
\end{table}
The calculation of $f_n$ for our simulation data is straightforward. Results corresponding to each amount of deposited layers, from 1 to 4, are included in Table~\ref{tab:overcharge}. Here is convenient to recall that these values correspond only to a deposition process under 0.1M of added salt conditions, since other discussed concentrations were used only for exposition of the films after their growth procedure, and no desorption of OE chains happend during such exposition. One can see that, for each amount of deposited layers but 3L, $f_n$ is clearly above the linear growth regime. Instead, for 3L it is significantly below. Fluctuations around the condition of linear growth regime for the first few deposited layers are frequently observed in both, experiments and simulations\cite{2013-ghostine, 2006-patel}.

\section{Conclusions}
We presented results of large scale fully atomistic molecular dynamics simulations of the layer-by-layer growth of an oligoelectrolyte multilayer, made by an alternating deposition of low molecular weight PDADMAC/PSS chains on a negatively charged silica substrate. Our simulation protocol mimics every step of the experimental layer-by-layer growth procedure. We focused our analysis on the comparison of the microscopic properties corresponding to systems with three (3L) and four (4L) deposited layers. This corresponds to an early, transient stage of the multilayer growth process that challenges the accuracy of experimental measurements, as the film thickness and the relative differences in the properties of consecutive layers at such stage are rather small. However, several properties measured in our simulations show at least qualitative agreement with experimental findings on early transient behaviors of analogous systems, in most cases corresponding to high molecular weight polyelectrolytes.

Our results indicate a slightly higher water uptake for the 3L PDADMAC ended system than for the 4L PSS ended one. The exposition to solvents with different ionic strengths ranging from 0 to 0.5M of added salt concentration has shown very little impact on this behavior. Only a weak indication of a decrease in the uptake with increasing ionic strength has been observed.

The intrinsic charge compensation mechanism in our multilayers is much stronger than the extrinsic one, independently from the amount of deposited layers and the background ionic strength. Comparatively, intrinsic compensation is slightly stronger in 3L systems than in 4L ones. The small extrinsic compensation that still exists is relatively more important for that type of oligoelectrolyte ending the multilayer.

Even though indirect experimental observations have predicted a low polymer entanglement and a strong internal layering for PDADMAC/PSS oligoelectrolyte multilayers, our simulated films show a rather high entanglement of the chains, that leads to an internal structure of numerous thin fuzzy layers. This might be a characteristic of early stages of multilayer formation in terms of both, the small number of deposited layers and the exclusive consideration of oligoelectrolytes that show considerably fast dynamically relaxation processes. 

Finally, for all but 3L systems we obtained overcharge fractions clearly above the linear regime condition. Since the experimental deposition conditions reproduced here correspond to linear growth, this result underlines the transient nature of the stage we are simulating.

Our results provide not only important insights in early stages of multilayer growth, but can serve also a reference for future multiscale models, that might be able to address the very slow dynamics processes that long polyelectrolytes would show and that certainly will take place during later stages of a multilayer growth.

\section*{Conflicts of interest}
There are no conflicts to declare.

\section*{Acknowledgements}
This research was partially supported by the Deutsche Forschungsgemeinschaft (DFG) within the Priority Program SPP 1369. P.A.S. acknowledges support from the Act 211 of the Government of the Russian Federation, contract No.~02.A03.21.0006. Simulations were carried out at bwGRiD \bibnote{BwGRiD (http://www.bw-grid.de), member of the German D-Grid initiative, funded by the Ministry for Education and Research (Bundesministerium fuer Bildung und Forschung) and the Ministry for Science, Research and Arts Baden-Wuerttemberg (Ministerium fuer Wissenschaft, Forschung und Kunst Baden- Wuerttemberg).} computing clusters and the Hermit Cray XE6 supercomputer of the High Performance Computing Center Stuttgart (HLRS).

\balance

\providecommand*{\mcitethebibliography}{\thebibliography}
\csname @ifundefined\endcsname{endmcitethebibliography}
{\let\endmcitethebibliography\endthebibliography}{}

\end{document}